\documentclass[epj,nopacs,fleqn]{svjour}
\pdfoutput=1

\usepackage{cite}
\usepackage{verbatim}
\usepackage{amsmath,amssymb}
\usepackage{color}
\usepackage{graphicx}
\usepackage{cancel}
\usepackage{url}
\usepackage{slashed}

\newcommand{\snu}{\tilde{\nu}}

\newcommand{\bra}{\mbox{\ensuremath{\mathcal{B}}}}
\newcommand{\mlsp}{m_{\tilde\nu_{\tau_1}}}
\newcommand{\mneut}{m_{\tilde{\chi}^0_1}}
\newcommand{\sigmav}{\langle \sigma v\rangle}

\renewcommand{\ell}{{l}}

\newcommand{\lambdabar}{{\hbox{$\lambda$\kern-1.ex\raise+0.45ex\hbox{--}}}}

\DeclareMathAlphabet{\mathpzc}{OT1}{pzc}{m}{it}

\begin{document}


\title{Monochromatic neutrino lines from sneutrino dark matter}

\author{
 Chiara Arina\inst{1},
 Suchita Kulkarni\inst{2},
 Joseph Silk\inst{1, 3,4,5}}

\institute{ 
Institut d'Astrophysique de Paris (UMR7095: CNRS \& UPMC- Sorbonne Universities), F-75014, Paris, France
  \and
 Institut f\"ur Hochenergiephysik,  \"Osterreichische Akademie der Wissenschaften, Nikolsdorfer Gasse 18, A-1050 Wien,  Austria
 \and
 AIM-Paris-Saclay, CEA/DSM/IRFU, CNRS, Univ. Paris VII, F-91191 Gif-sur-Yvette, France
\and
Department of Physics and Astronomy, The Johns Hopkins University Homewood Campus, Baltimore, MD 21218, USA
\and
BIPAC, Department of Physics, University of Oxford, Keble Road, Oxford
OX1 3RH, UK
}

\abstract{
We investigate the possibility of observing monochromatic neutrino lines originating from annihilation of dark matter. We analyse several astrophysical sources with overdensities of dark matter that can amplify the signal. As a case study, we consider mixed left and right handed sneutrino dark matter. We demonstrate that in the physically viable region of the model, one can obtain a prominent monochromatic neutrino line. We propose a search strategy to observe these neutrino lines in future generations of neutrino telescopes that is especially sensitive to dwarf spheroidal galaxies. We demonstrate that the  presence of massive black holes in the cores of dwarfs as well as of more massive galaxies substantially boosts any putative signal. In particular, dark matter in dwarf galaxies spiked by IMBH provides a  powerful means of probing low annihilation cross-sections well below $10^{-26} \rm cm^3 s^{-1}$ that are otherwise inaccessible by any  future direct detection or collider experiment.
}  

\date{}

\titlerunning{Neutrino lines from sneutrino DM}

\maketitle


\vspace*{-12cm}
\noindent
\small{HEPHY-PUB 954/15}
\vspace*{11cm}

\section{Introduction}\label{sec:intro}

Indirect detection of dark matter (DM) is an important probe of the nature of DM. It relies on the search for annihilation products of DM particles, which can include  photons, anti-matter and neutrinos. The usual strategy is to observe a secondary flux arising from DM annihilations, for instance at the center of the galaxy or in point sources. Despite the astrophysical uncertainties involved, limits from {\it e.g.} space and ground-based Cherenkov telescopes set important constraints on the thermal cross-section of the DM annihilations into gamma rays. In terms of the neutrino secondary flux, high energy neutrinos produced by the capture and annihilation of weakly interacting massive particles (WIMPs) in stellar cores are one of the smoking guns for DM detection~\cite{SOS, Gould:1987ir} and are actively being searched for by neutrino telescopes. Several studies, { \it e.g.}~\cite{Bergeron:2013lya,Enberg:2015qwa}, rely on this signature and help to determine future prospects for neutrino detection in supersymmetric (SUSY) models. 

Another search strategy for DM in astrophysical environments relies on spectral features, which arise in some DM scenarios, e.g. monochromatic lines that target the DM mass and serve as a smoking gun signature for WIMPs~\cite{Bouquet:1989sr}. Gamma-ray lines have been studied extensively (see this review for further references~\cite{Bringmann:2012ez}), while much less attention has been given to monochromatic neutrino lines~\cite{Barger:2007hj,Lindner:2010rr,Aisati:2015vma}, which however may have rather interesting features. 

In this paper, we study the prospects for detection of mono-chromatic neutrino lines from WIMP annihilations arising at tree level in certain DM scenarios. Neutrinos can be a particularly clean signature of DM annihilations as they propagate freely and undisturbed to the Earth, while the other particles undergo absorption and scattering processes before reaching the Earth. Indirect searches for DM are particularly sensitive to the astrophysical environment. This introduces several uncertainties in the limits derived using indirect detection techniques. However, we take advantage of this fact and assess the impact of enhancement in the indirect detection signatures due to DM overdensities in diverse astrophysical sources. Among alternative astrophysical environments with an enhanced DM density, we consider galaxies hosting a supermassive black hole (SMBH) and dwarf spheroidal galaxies (dSphs). We propose these astrophysical targets as a future detection strategy for DM particles at neutrino telescopes. We demonstrate that these searches can probe values of the annihilation cross-section well below $10^{-26} \rm cm^3 s^{-1}$. 

The presence of a SMBH might have dramatic consequences for DM annihilation. Such black holes (BHs) may be common signatures of galaxy formation, at least as envisaged in many scenarios of structure formation~\cite{2004MNRAS.354..427I,Rashkov:2013uua}, and are capable of boosting the DM density within its sphere of gravitational influence~\cite{Gondolo:1999ef}. In particular BHs can act as a particle accelerator~\cite{Banados:2009pr}, especially in the case of Kerr BHs. For WIMPs close to the BH horizon, the scattering may occur at almost infinite center-of-mass energies~\cite{Schnittman:2014zsa, Berti:2014lva, Zaslavskii:2014jea}, for example boosting cross-sections which are p-wave and usually suppressed at the present epoch~\cite{Shelton:2015aqa}, or exciting resonances and producing a forest of gamma-ray lines~\cite{Cannoni:2012rv,Arina:2014fna}. In our case, the dominant cross-section for producing neutrino lines however is velocity-independent and we will not investigate velocity-dependent enhancements here. 

We focus here on the more generic scenario in which the SMBH may increase the DM density distribution towards the center of the galaxy and form a so-called DM spike~\cite{Gondolo:1999ef}. In the standard picture, the DM density profile follows a power law, $\rho \propto r^{-\gamma}$ in the inner region, with $r$ the distance from the galactic center and $\gamma \sim 1$ the slope for a Navarro-Frenk-White (NFW) profile~\cite{Navarro:1995iw}. However, very close to the BH, the DM density profile may rise very steeply. Assuming that the BH grew adiabatically, $\gamma$ is expected to instead lie between 2.25 and 2.5, leading typically to $ \gamma \equiv \gamma_{\mathrm{sp}} = 7/3$ in the very inner region. Such an enhancement of the DM density profile is referred to as a spike and enhances the luminosity of the flux originating from annihilating WIMPs. 

The existence of such a DM spike in the vicinity of SMBHs is still under debate. If the BH growth was not adiabatic, then the inner DM density profile would behave instead as $\rho \propto r^{-4/3}$~\cite{Ullio:2001fb,Gnedin:2004cx}. Besides, if the DM halo itself underwent a merger, or if the BH did not grow exactly at the center of the DM halo, the inner DM halo profile would follow $\rho \propto r^{-1/2}$, thus considerably reducing the flux expected from WIMPs. Even if a spike could form with $\gamma_{\rm{sp}} \sim 7/3$, the process of dynamical relaxation by WIMP scattering off stars could partially smooth the spike and lead to a much shallower slope, $\rho \propto r^{-3/2}$, which would coincidentally correspond to a Moore profile~\cite{Moore:1999gc}. Constraints from a DM spike in terms of annihilating DM producing an electromagnetic flux have been discussed in~\cite{Gorchtein:2010xa,Belikov:2013nca,Lacroix:2015lxa}. Use of a  secondary neutrino flux to probe the DM spike has been initially proposed in~\cite{Gondolo:1999ef}, however monochromatic neutrino lines have not yet been discussed in this context.

Another interesting target for  indirect detection of DM is dwarf spheroidal galaxies (dSphs) in the Milky Way Galaxy (MWG)~\cite{Evans:2003sc}, along with the Galactic Center. What makes the former really interesting is their low astrophysical background, high mass-to-light ratios and proximity to the Earth (see for example~\cite{Conrad:2015bsa}). To further enhance signals from WIMP annihilations, we study the impact of a DM minispike due to an intermediate massive black hole (IMBH) hosted in the inner region of nearby dSphs~\cite{Bertone:2005xz}. Although the presence of the DM minispike is also a matter of debate and the answer is still unclear, it should be kept in mind that DM scattering off stars will generally be much less efficient than in supergiant galaxies, as there are far fewer stars, if any, in the cores of the ultra faint dSphs that are heavily DM-dominated. Indeed there is increasing observational evidence that dwarfs contain central massive black holes. The AGN fraction may be as large as $\sim 10\%$ for the lowest mass nearby infrared-detected dwarfs and $\sim 1\%$ if optically selected \cite{2014ApJ...784..113S} with active AGN found for black holes in the mass range $10^2-10^6 \rm M_\odot$ \cite{2014arXiv1411.3844M}. X-ray selected dwarfs are rarer, with central massive  BH candidates amounting to $\sim 0.1\%$\cite{2015ApJ...805...12L}.  Some degree of theoretical plausibility is added, in addition to the hierarchical structure arguments cited previously, by the suggestion that IMBHs in dwarfs are a crucial element of early feedback, in terms of ejecting baryons, and reducing the baryon content of the host galaxy~\cite{2012MNRAS.427.2625P}, an outstanding problem for MWG-like galaxies where the missing baryons are presumed to have been heated and ejected beyond the virial radius by unknown processes~\cite{2015arXiv150603469B}. Constraints on minispikes have been studied in the case of WIMPs primarily producing a secondary gamma ray flux~\cite{Gonzalez-Morales:2014eaa,Wanders:2014xia}, however dSphs in general have been poorly exploited for DM searches by neutrino detectors~\cite{Aartsen:2013dxa}. 

Remarkably, indirect detection of decaying DM via neutrino detection becomes compatible with gamma ray signals for  DM masses above $\sim 1\rm\,  TeV$~\cite{Aisati:2015vma}, and might be relevant for heavy DM candidates as in~\cite{Arina:2014fna}. One can however expect that future improvements to neutrino telescopes, for example with more closely spaced strings similarly to an enlarged  DeepCore-like detector~\cite{Aartsen:2015xej}, will improve sensitivity considerably  down to $\sim \rm 100\,  GeV$.  Here, we explore a SUSY thermal DM scenario that predicts enhanced branching ratios to monochromatic $\nu$ signals at these energies and illustrate the detection capabilities of future neutrino telescopes.

The focus of this paper is to analyse neutrino lines originating from annihilating DM. As a concrete example of DM annihilation to neutrino final states, we take a supersymmetric scenario. The model consists of the Minimal Supersymmetric Model (MSSM) augmented by a RH neutrino superfield, which provides a mostly right-handed (RH) mixed sneutrino as the lightest supersymmetric particle (LSP)~\cite{Borzumati:2000mc,Arkani-Hamed:2000bq}. This model is well motivated as it addresses two basic problems: the origin of neutrino masses and the nature of DM. Purely left-handed (LH) sneutrinos as given by the MSSM have been studied in~\cite{Ibanez:1983kw,Hagelin:1984wv} as DM candidates, however due to their large coupling to the $Z$ boson, they are excluded from direct detection searches as a dominant component of DM. By introducing RH chiral superfields, the LH sneutrinos can mix with its RH counterparts and lead to the required relic density~\cite{Planck:2015xua} while fulfilling the direct search exclusion bounds. On the other hand, the RH fermionic field, a RH neutrino, provides the mechanism to generate neutrino masses and a connection between DM and neutrino phenomenology. 

Several possibilities for  neutrino mass mechanisms and their consequences for sneutrino DM phenomenology have been investigated, for instance Dirac masses~\cite{Arkani-Hamed:2000bq,Arina:2007tm,Belanger:2010cd,Dumont:2012ee}, seesaw mechanisms~\cite{Grossman:1997is,Hall:1997ah,Dedes:2007ef,Arina:2007tm} and inverse seesaw mechanisms~\cite{Arina:2008bb,DeRomeri:2012qd}. In this paper, we consider the simplest possibility of Dirac masses, {\it e.g.} without lepton-violating terms, called MSSM+RN as in~\cite{Arina:2013zca,Arina:2015uea}. Neutrino lines produced by various neutrino mass generation mechanisms have been analysed in~\cite{Lindner:2010rr} in an effective field theory approach without however enforcing the DM constraints. 

In our case, the sneutrino is the dominant DM component and compatible with the exclusion bound by LUX~\cite{Akerib:2013tjd} for direct searches. We show that sneutrino pair annihilations produces a sharp and enhanced monochromatic neutrino line at tree level, besides the usual secondary neutrino flux. As will be explained below, apart from the mass and mixing angle of the lightest sneutrino, the monochromatic neutrino line is mainly sensitive to the mass spectrum and nature of  the neutralino. Thus, it provides an excellent complementarity to collider searches, in case the rest of the MSSM+RN spectrum is beyond the reach of the LHC.  Other indirect detection prospects for sneutrinos have been studied in~\cite{Hooper:2004dc,Arina:2007tm,Dumont:2012ee}, while Refs.~\cite{Arina:2007tm,Allahverdi:2009se,MarchRussell:2009aq,Belanger:2010cd} present a detailed analysis of the neutrino flux or line from the Sun or from the Earth. In general, the MSSM+RN LHC signatures can be quite distinct from those of the conventional neutralino LSP, as has been discussed in~\cite{Thomas:2007bu,Belanger:2011ny,BhupalDev:2012ru,Guo:2013asa,Harland-Lang:2013wxa,Arina:2013zca} and in~\cite{Arina:2015uea} in the framework of simplified model spectra (SMS). From this latter analysis, which takes into account an extensive sampling of the MSSM+RN parameter space, it has emerged that there are almost always parameter combinations such that the limits from SMS can be avoided. It was shown that the direct detection experiments provide an excellent complementarity to the LHC searches for thermal sneutrino DM~\cite{Arina:2015uea}. 

At a practical level, we consider a subset of points analysed in~\cite{Arina:2015uea}. The MSSM+RN have 13 free parameters (gaugino masses, scalar masses for LH sleptons and squarks, RH sneutrinos, RH charged sleptons and squarks, trilinear couplings for all the scalar sector and finally the ordinary parameters describing the Higgs sector) whose input value is given at the Grand Unification scale $\sim 10^{16}$ GeV. The MSSM+RN SUSY particle spectrum is computed with the code~\texttt{SoftSusy}, appropriately modified to adapt to \texttt{micrOMEGAS\_3.6}~\cite{Belanger:2013oya} and \texttt{micrOMEGAS\_4.1}~\cite{Belanger:2014vza} for relic density and particle differential spectra computations. We use a nested sampling algorithm \texttt{MultiNest\_v3.2}~\cite{Feroz:2007kg,Feroz:2008xx} to sample efficiently these free parameters. In the model likelihood calculations, we include all DM constraints: the sneutrino satisfies relic density constraints from Planck measurements~\cite{Planck:2015xua} and the elastic scattering cross-section off nuclei is compatible with the 90\% confidence level (CL) of the LUX exclusion bound~\cite{Akerib:2013tjd}. All details about the sampling procedure, the constraints and measurements implemented in the likelihood function, and the LHC phenomenology are provided in~\cite{Arina:2015uea}. 

The rest of the paper is organised as follows. In the next section we discuss the modelling of the DM density spikes in SMBHs, the dwarf spheroidal galaxies and the DM minispike. In Sec.~\ref{sec:param} we study a simplified model for the monochromatic neutrino line to extract the relevant dependence on the theoretical parameters.  Section~\ref{sec:set} describes briefly how neutrino flavor oscillations modify the line signal. More importantly, it states the setup for an ideal neutrino telescope  capable of detecting the monoenergetic neutrino lines in the $\mathcal{O}(100)$ GeV energy range. In Sec.~\ref{sec:res} we provide details of the MSSM+RN model in which the sneutrino is a successful DM candidate and discuss the expected flux of monochromatic neutrino lines. We give our conclusions in Sec.~\ref{sec:concl}. 
%
\section{Neutrino flux from WIMP annihilation in astrophysical sources with high DM density}\label{sec:astro}

Sneutrinos as DM candidates can annihilate with a small but non negligible probability into the Standard Model (SM) particles producing several possible final states, along with neutrinos. The expected neutrino differential flux in an astrophysical source is given by:
\begin{equation}
\frac{{\rm d} \Phi_\nu}{{\rm d} E} = \frac{1}{8 \pi}  \, \xi^2 \,  \frac{\sigmav}{ m^2_{\snu_{\tau_1}} }\, \frac{{\rm d} N_{\nu}}{{\rm d} E} \, \Phi_{\rm Astro}\,.
\label{eq:dflux}
\end{equation}
The first part of Eq.~\ref{eq:dflux} depends only on the WIMP model. In particular $\xi \equiv \Omega h^2_{\rm \snu}/\Omega h^2_{\rm Planck}$ (relevant for subdominant dark matter), $\sigmav$ is the total annihilation cross-section of sneutrino pairs at the present epoch, $m^2_{\snu_{\tau_1}}$ is the sneutrino mass and ${\rm d} N_{\nu}/{\rm d} E$ the differential neutrino spectrum per annihilation event, as a function of  energy $E$:
\begin{equation}
\frac{{\rm d} N_{\nu}}{{\rm  d} E} = \left\{
\begin{array}{l c}
\bra_{\nu}^\tau\,  \frac{{\rm d} N_{\nu_{\rm line}}}{{\rm d}E}\, \delta(E - m_{\snu}) & {\rm line}\,,\\
\sum_k \bra_{\nu}^k\,  \frac{{\rm d} N_{\nu_k}}{{\rm d} E}& {\rm secondary} \,.
\end{array}
\right. 
\end{equation}
$\bra_{\nu}^j$ is the fractional contribution into neutrinos of each channel $k$ and ${\rm d} N_{\nu_j}/{\rm d}E$ is the number of neutrinos produced in the $j$th channel. \footnote{The sneutrinos are not self-conjugate particles, hence the secondary neutrino differential spectrum in Eq.~\ref{eq:dflux} should be rescaled by an additional factor $1/2$.} For processes dominated by s-wave, see Sec.~\ref{sec:param}, the particle physics part of Eq.~\ref{eq:dflux} can be computed independently of the astrophysical factor, $\Phi_{\rm Astro}$, which we discuss below.

\begin{table*}[t!]
\caption{Parameters relevant for the analysis for each SMBH we consider. In all cases we assume $t_{BH} = 10^{10}$ years, while black hole mass $M_{BH}$ and distance $D$ is taken from ~\cite{Gorchtein:2010xa,Lacroix:2015lxa}.\label{tab:BHs}}
\begin{center}
\begin{tabular}{r|cccc|c|c}
\hline
SMBH & $M_{\rm BH}$ [$M_{\odot}$] & $R_{\mathrm{S}}$ [pc] & $D$ [Mpc] & $\rho_0$ [GeV cm$^{-3}$] & $\Phi_{\rm Astro}$ [GeV$^2$ cm$^{-5}$] & Declination \\
\hline
M87 &  6.4 $\times 10^{9}$ & $6.1 \times 10^{-4}$  & 16.4 & 2.3 & $3.5 \times 10^{11}$ & +12$^\circ$\\
CenA & 5.5 $\times 10^{7}$ & $5.3 \times 10^{-6}$ & 3 & 9 $\times 10^{5}$ & $4.3 \times 10^{20}$ & -43$^\circ$\\
NGC1277 & 1.7 $\times 10^{10}$ & $1.6 \times 10^{-3}$ & 20 & 495 & $2.5 \times 10^{12}$ & +41$^\circ$\\
\hline
\end{tabular}
\end{center}
\end{table*}
\subsection{Dark matter spike in supermassive black holes}\label{sec:spike}
We begin with supermasssive black holes in massive galaxies. The MWG SMBH is not the optimal target, mostly because the presence of a nuclear star cluster guarantees that scattering will soften any DM spike, the scattering time-scale being of order of a Gyr~\cite{Vasiliev:2008uz}. Of course continuing growth of the BH may partially compensate this, but the detailed evolution is  complicated by the competition between growth of the nuclear star cluster by both gas accretion and in situ star formation as well as infall of globular star clusters~\cite{Antonini:2015sza}. A much better target is a SMBH in a massive spheroidal galaxy such as M87 or Centaurus A (CenA). In the former case the stellar heating time of any putative DM spike is of order 10$^{14}$ Gyr, and therefore presents the possibility of an excellent DM signal amplifier~\cite{Lacroix:2015lxa}.  

We then consider three optimised galaxy targets, known to host a SMBH in their core and we assume that a DM spike has formed and survived all possible disruption processes. For instance, we analyse M87, a supergiant elliptical galaxy in the constellation Virgo at a distance of 16 Mpc from the Sun. In M87, it is plausible that the scattering off stars has been inefficient in erasing the spike because it is dynamically young. The relaxation time for M87 is estimated to be $t_{\mathrm{r}} \sim \mathcal{O}(10^{5})$ Gyr due to the strong dependence on the velocity dispersion of the stars and the relatively low core density.\footnote{Dynamical heating by stars is inefficient when the dynamical relaxation time $t_{\mathrm{r}}$ in the BH core is larger than the Hubble time ($\sim10^{10}\ \rm yr$).} Hence, a spike formed at early times is much more likely to have survived galaxy dynamics up to the present epoch in M87 than in the Milky Way, which has a relaxation time of $\sim 2.5$ Gyr. Similar reasoning holds for NGC1277 and CenA. We consider NGC1277, which is a lenticular galaxy in the constellation Perseus, as it may host one of the largest SMBH ever measured~\cite{ngc1277nature,Emsellem:2013gya} and is at a distance similar to M87 with respect to us. Finally, CenA is in the constellation Centaurus and despite of hosting a smaller SMBH compared to M87 and NGC1277, it has the advantage of being one of the closest radio galaxies to the Sun. All relevant details about these three galaxies and their SMBHs are provided  in Tab.~\ref{tab:BHs}.

The dark matter density spike for a SMBH consists of two components, $\rho_{pl}$ the DM density of the plateau and $\rho_{sp}$ the spike profile. With the assumption of adiabatic growth of the DM spike, its density is given by~\cite{Gondolo:1999ef}:
\begin{equation}
\label{eq:rhor}
\rho_{\rm BH} (r) = \left\{
\begin{array}{l c}
 0  & r < 4 R_{\mathrm{S}} \\
 \frac{\rho_{\mathrm{sp}}(r) \rho_{\mathrm{pl}}}{\rho_{\mathrm{sp}}(r) + \rho_{\mathrm{pl}}}  & 4 R_{\mathrm{S}} \leq r < R_{\mathrm{sp}} \\
 \rho_{0} \left( \frac{r}{r_0} \right)^{-\gamma} \left( 1 + \frac{r}{r_0} \right) ^{-2} &  r \geq R_{\mathrm{sp}},
\end{array}
\right.
\end{equation}
with $R_{\mathrm{sp}}$ being the spike radius. This expression relies on the assumption that the spike has grown from a DM density profile $\propto \rho_{0} \left( r/r_0 \right)^{-\gamma}$. The plateau density depends on the DM annihilation rate as:
\begin{equation}\label{eq:rhopl}
\rho_{\mathrm{pl}} = \frac{m_{\tilde{\nu}}}{\sigmav t_{\mathrm{BH}}},
\end{equation}
with $t_{\mathrm{BH}}$ the age of the BH with mass $M_{\rm BH}$. The spike profile is
\begin{equation}\label{eq:rhopsike}
\rho_{\mathrm{sp}}(r) = \rho_{\mathrm{R}} g_{\gamma}(r) \left( \frac{R_{\mathrm{sp}}}{r} \right)^{\gamma_{\mathrm{sp}}},
\end{equation}
with
\begin{eqnarray}
& \rho_{\mathrm{R}} = \rho_{0} \left( \frac{R_{\mathrm{sp}}}{r_0} \right) ^{-\gamma}\,,\label{eq:rhospike2}\\
& g_{\gamma}(r) \approx \left( 1 - \frac{4 R_{\mathrm{S}}}{r} \right) ^{3}\,,\\
& R_{\mathrm{sp}} = \alpha_{\gamma} r_0 \left( \frac{M_{\mathrm{BH}}}{\rho_{0}r_0^{3}} \right) ^{\frac{1}{3-\gamma}}
\end{eqnarray}
and $\gamma_{\mathrm{sp}} = \frac{9-2\gamma}{4-\gamma}$. For all numerical evaluations we set $r_0 = 20\ \rm kpc$ similar to the Milky Way, $\alpha_{\gamma} = 0.1$ and $\gamma = 1$ (which correspond to a NFW density profile and gives a spike slope $\gamma_{\mathrm{sp}} = 7/3$). The normalization $\rho_{0}$ is fixed according to the prescriptions in~\cite{Gorchtein:2010xa} and is in agreement with the values reported in~\cite{Lacroix:2015lxa}.

Considering the galaxies hosting the SMBH as point sources, the astrophysical flux is given by:
\begin{equation}
\Phi_{\rm Astro} = N_{\rm esc}\,  \frac{1}{D^2} \, \int_{4 R_{\mathrm{S}}}^{\infty} {\rm d}r \, r^2\,  \rho_{\rm BH}^2(r) \,,
\end{equation}
where $D$ and $R_S$ are respectively the distance to the Earth and the Schwarzschild radius of the SMBH. $N_{\rm esc}$ parametrises the fraction of neutrinos produced by WIMP annihilations that can actually escape from the BH. Numerical simulations tend to indicate that $N_{\rm esc}$ is always larger than 90\% even close to the event horizon for a Kerr BH~\cite{Schnittman:2015oma}, a value we take as a reference for the numerical analysis.

The behavior of $\rho_{\rm BH}(r)$ is shown in Fig.~\ref{fig:spikesdwarf} top panel, for a DM mass of 100 GeV and two different values of the total thermally-averaged annihilation cross-section. For a cross-section close to the value of a standard thermal relic ({\it i.e.} $\sim 10^{-26}$ cm$^3$/s), the plateau region is clearly visible, while if $\sigmav \leq 10^{-30}$ cm$^3$/s the DM interaction is not strong enough to flatten out  the spike for NGC1277 and M87. This means that below a certain value of $\sigmav$, it is possible to probe the entire DM spike leading to a maximal enhancement in the neutrino flux, independent of the particle physics. This is not true for CenA, where the DM density is larger. In this case, even small cross-section values of 
$10^{-30}$ cm$^3$/s flatten out the DM spike. We can approximate the DM density profile as:
\begin{equation}
\rho_{\rm BH} (r) = \left\{
\begin{array}{l c}
\rho_{\mathrm{pl}}  & {\rm if} \, \, \,  \rho_{\mathrm{sp}}(r) >> \rho_{\mathrm{pl}}\,,\\
\rho_{\mathrm{sp}}(r) &  {\rm if} \, \, \,  \rho_{\mathrm{sp}}(r) << \rho_{\mathrm{pl}}\,.
\end{array}
\right.
\end{equation}
In the former case, the astrophysical flux is proportional to the plateau density times $R_{\mathrm{S}}^3/D^2$, while in the second case the dependence is $\Phi_{\rm Astro} \sim \rho_0^2 r_0^2 (\mathcal{O}(R_{\mathrm{S}}) + \mathcal{O}(R_{\mathrm{sp}}))$. Hence, in the first case, the luminosity gets enhanced for very massive even though distant SMBHs, while in the latter case this is not true anymore. This will be a relevant issue for our study, as discussed in Sec.~\ref{sec:res}.

\subsection{Dwarf spheroidal galaxies}\label{sec:dwarf}
Let us now turn our attention to $\Phi_{\rm Astro}$ from dSphs. The expected flux luminosity from dSphs is given by the so-called J-factor~\cite{Bergstrom:1997fj}:
\begin{equation}\label{eq:jfactor}
\Phi_{\rm Astro} \equiv J(\Delta \Omega)= \int_{\Delta \Omega} {\rm d}\Omega' \int_{los} \rho_{\rm dwarf}^2(r(s,\theta))\,{\rm d}s\,,
\end{equation}
with $s$ being the distance along the line of sight and $\theta$ the opening angle of the cone. For the J-factor, we use the values computed in~\cite{Bonnivard:2015xpq,Bonnivard:2015tta} for integrating over a $1^{\circ}$ angle. This allows  a better assessment of the various sources of uncertainties that affect the J-factor.\footnote{Uncertainties include for instance the choice of the DM density profile, the sample size and the quality of kinematic data; for a thorough discussion we refer to~\cite{Bonnivard:2015xpq}.}

Numerical simulations suggest  that IMBHs with masses  between $10^2 M_\odot - 10^4 M_\odot$ may have grown and formed a spike in up to 10\% of the cases for Draco-like dSphs~\cite{VanWassenhove:2010bj}, and this will be our underlying working hypothesis. To model this minispike hosted at the center of the dSph, we use a similar prescription for $\rho_{\rm dwarf}(r)$ as for the SMBHs except for a modification in Eq.~\ref{eq:rhor}: $\rho_0$ is now defined as the normalization for NFW density profile and $r_0$ is the scaling radius as usual. The modeling of $\rho_{\rm sp}(r)$ is given by Eq.~\ref{eq:rhospike2}, with $R_{\rm sp}$ and  $\rho_{\rm R}$ obtained with the prescriptions in~\cite{Bertone:2005xz,Wanders:2014xia}.
\begin{table}[t!]
\caption{Distance and J-factors ($1^{\circ}$) for each dSPh we consider in the analysis, from~\cite{Bonnivard:2015xpq,Bonnivard:2015tta}.\label{tab:dwarf}}
\begin{center}
\begin{tabular}{r|c|c}
\hline
dSph &  $D$ [kpc] &  $J(1^{\circ})$ [GeV$^2$ cm$^{-5}$]  \\
\hline 
\multicolumn{3}{c}{Northern sky}\\
\hline
Draco &   80 & $2.11 \times 10^{19}$ \\
Ursa Minor & 66 & $1.24 \times 10^{19}$ \\
Sextans & 86 & $8.09 \times 10^{17}$ \\
Leo I & 250 & $8.87 \times 10^{17}$ \\
Leo II & 205 & $1.37 \times 10^{18}$ \\
\hline
\multicolumn{3}{c}{Northern sky (ultra faint)}\\
\hline
Segue I & 23 & $2.06 \times 10^{17}$ \\
Ursa Major II & 30 & $1.87 \times 10^{20}$ \\
Segue II & 35 & $1.72 \times 10^{19}$ \\
Willman I & 38 & $4.75 \times 10^{19}$ \\
Coma & 44 & $8.32 \times 10^{19}$ \\
Bo\"{o}tes I & 66 & $6.07 \times 10^{18}$ \\
Ursa major I & 97 & $6.79 \times 10^{18}$ \\ 
Hercules & 132 & $1.99 \times 10^{18}$ \\
Canis Venatici II & 160 & $4.13 \times 10^{17}$ \\
Canis Venatici I & 218 & $4.50 \times 10^{18}$ \\
Leo V & 180 & $1.88 \times 10^{16}$ \\
LeoT & 407 & $4.80 \times 10^{17}$ \\
\hline
\multicolumn{3}{c}{Southern sky}\\
\hline
Carina & 101 & $1.05 \times 10^{18}$ \\
Fornax &138 & $7.07 \times 10^{17}$  \\
Sculptor & 79 & $4.30 \times 10^{18}$ \\
\hline
\multicolumn{3}{c}{Southern sky (ultra faint)}\\
\hline
Leo IV & 160 & $ 2.14 \times 10^{16}$ \\
Reticulum II & 30 & $5.88 \times 10^{20}$ \\
\hline
\end{tabular}
\end{center}
\end{table}
\begin{table*}[t!]
\caption{Minispike parameters, $\rho_0$ and scaling radius for Draco and Reticulum II for two assumed BH masses. $r_0$ and $\rho_0$ for Draco are taken from~\cite{Wanders:2014xia} while for ReticulumII we use~\cite{Geringer-Sameth:2014yza,Bonnivard:2015tta}. In all cases we assume $t_{\rm BH} = 10^{10}$ years.\label{tab:dwarf2}}
\begin{center}
\begin{tabular}{r|r|cccc}
\hline
& dSph &  $r_0$ [kpc] &  $\rho_0$ [GeV cm$^{-3}$]  & $R_{sp}$ [pc] & $\rho_{\rm R}$ [GeV cm$^{-3}$]   \\
\hline
$M_{\rm BH} = 10^4 (10^2) M_{\odot}$ & Draco &  2.09  & $0.99$ & 1.5 (0.15) & $1.3 \times 10^3 (1.3 \times 10^4)$  \\
$R_S = 9.57 \times 10^{-9} ( \times 10^{-12})$ pc & Reticulum II & 4.28 & $ 2.81$&  0.63 (0.063) & $1.9 \times 10^4 (1.9 \times 10^5)$  \\
\hline
\end{tabular}
\end{center}
\end{table*}
\begin{figure}[t]
\centering
\includegraphics[width=1.\columnwidth,trim=0mm 10mm 0mm 15mm, clip]{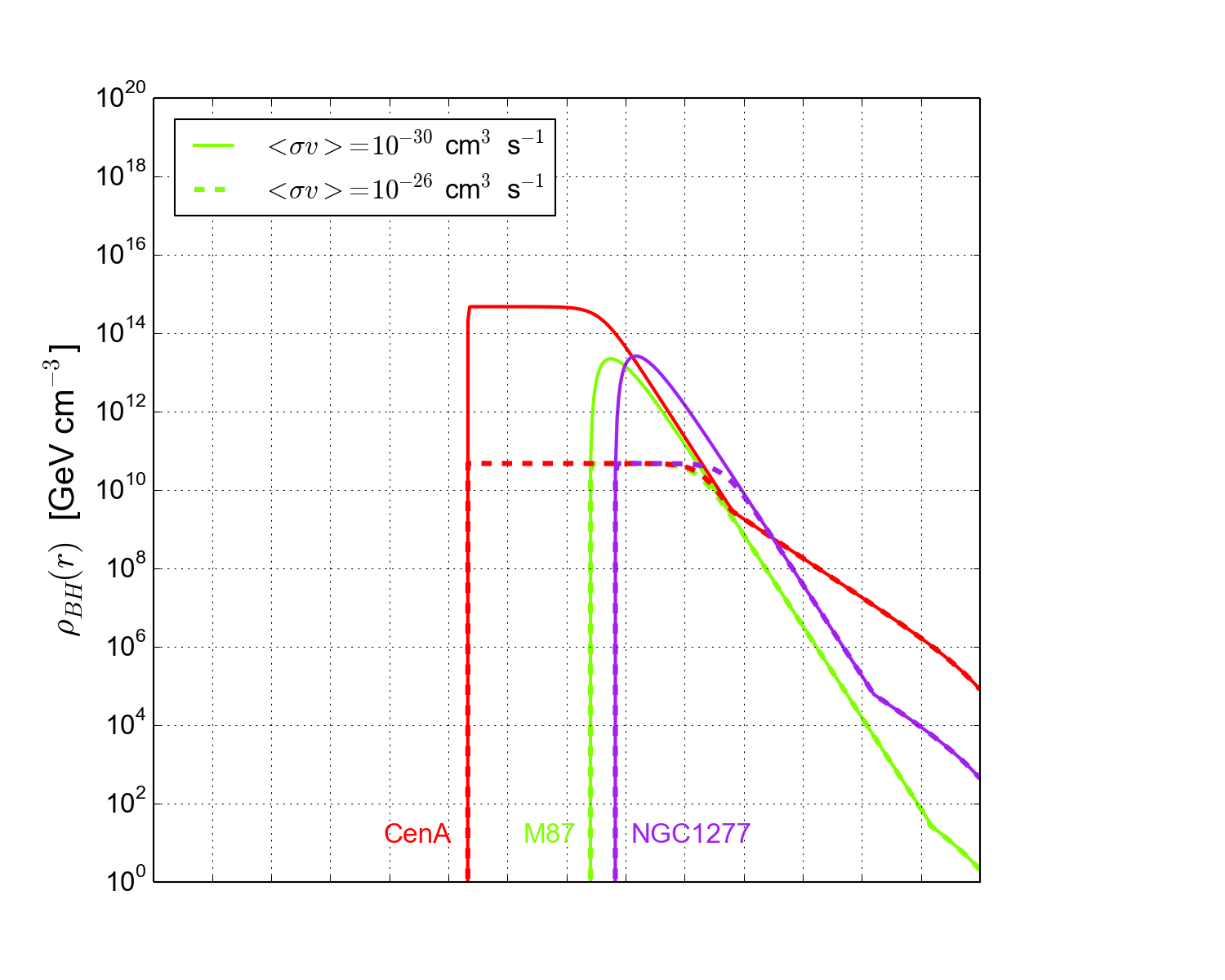}
\\
\centering
\includegraphics[width=1.\columnwidth,trim=0mm 0mm 0mm 15mm, clip]{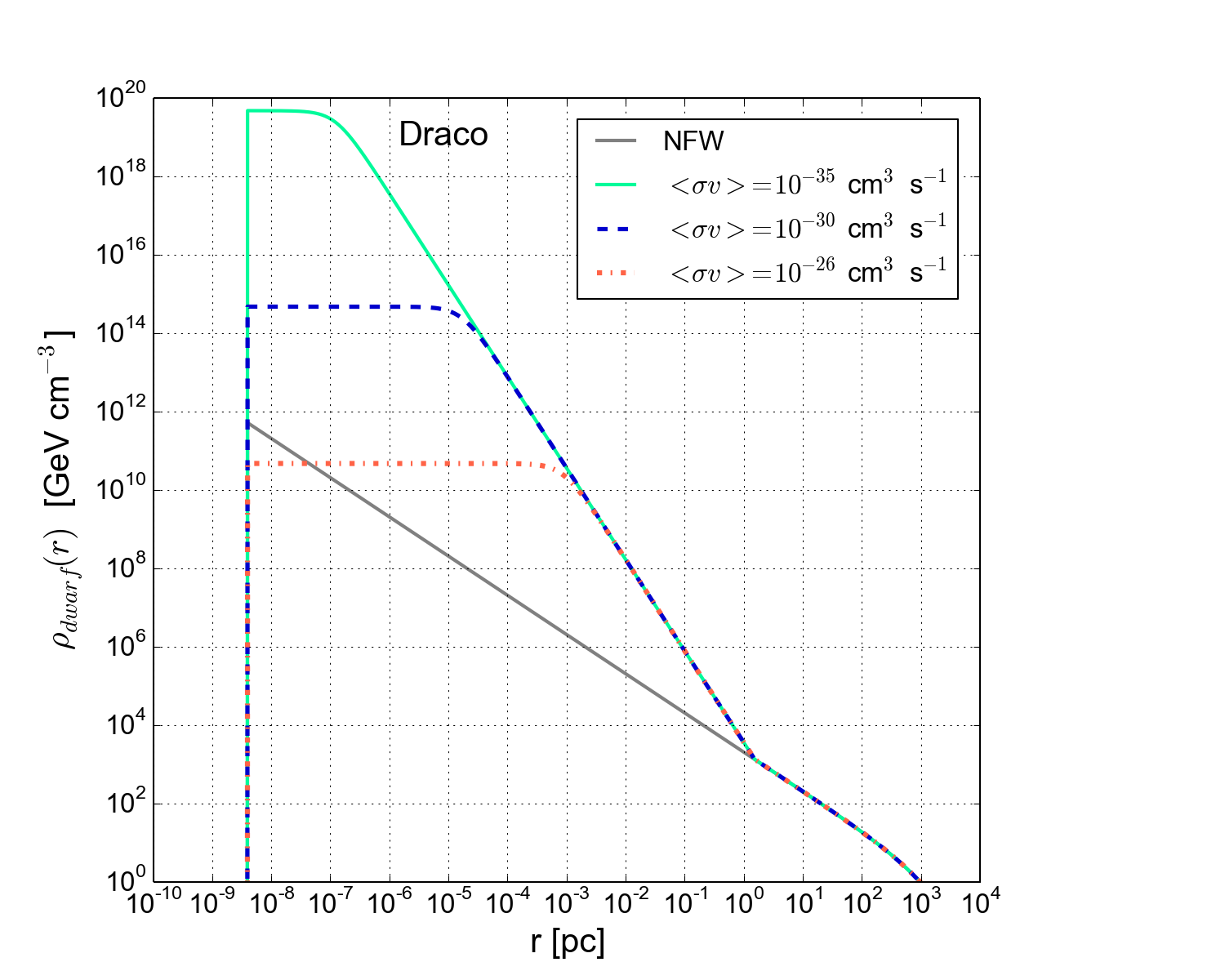}
\caption{{\it Top:} The DM density profile distribution for the three SMHBs is shown as a function of the distance from the center, as labelled. {\it Bottom:} Same as top panel but for a BH mass of $10^4 M_\odot$ in the  Draco dSph. The NFW density profile is denoted by the solid gray line for reference. In both panels we show the density plateau for different values for the thermal averaged cross-section as labelled, for a fixed DM mass of 100 GeV. 
\label{fig:spikesdwarf}}
\end{figure}

While about 25 nearby dSphs were known as of early 2015, recently imaging data from the Dark Energy Survey has led to the discovery of nine new Milky Way satellites in the Southern sky~\cite{ninedwarfs,desdwarf}, increasing the total number to 34. Among all these dSphs, we consider the 8 galactic dSPhs that have the best measured stellar kinematics, and 14 ultra faint galactic dSPhs, for which the kinematical data are more uncertain but however are interesting due to their proximity and sky coverage. The relevant numbers are reported in Tab.~\ref{tab:dwarf} and the separation into Northern and Southern hemispheres will be useful for assessing the sensitivity of neutrino telescopes. 

Table~\ref{tab:dwarf2} lists all relevant parameters for the two dSphs we consider in detail, Draco in the Northern hemisphere and Reticulum II in the Southern sky, for two different IMBH masses, $10^2 M_\odot$ and $10^4 M_\odot$. The total luminosity is given by the flux along the line of sight from the minispike plus the standard J-factor, Eq.~\ref{eq:jfactor}. The choice of these two dSPhs is dictated by the robustness of the J-factor against uncertainties~\cite{Bonnivard:2015xpq,Bonnivard:2015tta}. Reticulum II, in addition to being robust against uncertainties has one the largest J-factors among all dSPhs listed in Tab.~\ref{tab:dwarf}. In the Northern sky, the largest J-factors are provided by  Ursa Major II and Willman I; however these ultra faint galaxies possess either a velocity field indicating strong tidal disruption effects from the Milky Way or nonequilibrium kinematics. Given its proximity and large DM content, Reticulum II seems to be one of the most promising objects for WIMP annihilation searches.

Figure~\ref{fig:spikesdwarf} shows the behavior of the DM density profile as soon as the DM spike sets in for Draco, in the bottom panel. The DM density is enhanced by a few orders of magnitude with respect to the case of the SMBHs (top panel). As opposed to the case of the SMBHs, small annihilation cross-sections $\mathcal{O}(10^{-35})$ cm$^3$ s$^{-1}$ are not able to probe the full spike, as clearly indicated by the presence of the plateau due to WIMP annihilation. The change in the BH mass in dSphs does not significantly modify the picture: a change from $M_{\rm BH} = 10^4 M_\odot$ to $M_{\rm BH} = 10^2 M_\odot$ changes the cutoff at $4 R_s$ by a factor of $10^3$. The overall larger DM density and the vicinity of the dSphs with DM minispikes partially compensates the smallness of $R_S$ in $\Phi_{\rm astro} \sim R_S^3/D^2$.  Thus we expect to have a larger neutrino flux from the dSph + minispike with respect to the SMBHs.

It  is clear that observations of dSphs with minispikes provide a more promising way to observe monochromatic neutrino lines than SMBH. There is of course an additional layer of uncertainty: SMBHS are known to be present in massive galaxies, whereas the occurrence of  IMBHs in dwarfs or ultra faint dwarfs is  dependent on theoretically-motivated  but plausible speculation, with some modest observational support for the presence of IMBHs in the more  massive dwarfs. Having completed our discussion about the astrophysical part of Eq.~\ref{eq:dflux}, we now turn our attention to the particle physics model under consideration and sketch the conditions necessary to get the monochromatic neutrino lines from WIMP annihilation.

\section{Analysis of the monochromatic neutrino line}\label{sec:param}

\subsection{MSSM+RN model}\label{sec:modeltrue}

The model we consider is the MSSM with the inclusion of the RH neutrino superfield (MSSM+RN) as defined in~\cite{Arkani-Hamed:2000bq,Borzumati:2000mc,Arina:2007tm}. We briefly illustrate the main features of the model for clarity. The superpotential for Dirac RH superfields is given by
\begin{equation}\label{lrsuppot}
W = \epsilon_{ij} (\mu \hat H^{u}_{i} \hat H^{d}_{j} - Y_{l}^{IJ} \hat H^{d}_{i} \hat L^{I}_{j} \hat R^{J}
+ Y_{\nu}^{IJ} \hat H^{u}_{i} \hat L^{I}_{j} \hat N^{J} )\,,
\end{equation}
where $Y_{\nu}^{IJ}$ is a real and diagonal matrix in flavor space, from which the neutrino mass is $m_D^{I} = v_{u}Y_{\nu}^{II}$. 
It should be noted that lepton-number violating terms are absent in this framework and their inclusion is left for future work. 
The RH scalar fields add new terms in the soft-breaking potential 
\begin{eqnarray}\label{lrsoftpot}
& V_{\rm soft} = (M_{L}^{2})^{IJ} \, \tilde L_{i}^{I \ast} \tilde L_{i}^{J} + 
(M_{N}^{2})^{IJ} \, \tilde N^{I \ast} \tilde N^{J} - \nonumber\\
&  [\epsilon_{ij}(\Lambda_{l}^{IJ} H^{d}_{i} \tilde L^{I}_{j} \tilde R^{J} + 
\Lambda_{\nu}^{IJ} H^{u}_{i} \tilde L^{I}_{j} \tilde N^{J})  + \mbox{h.c.}]\,,
\end{eqnarray}
where the matrices $M_{N}^{2}$ and $\Lambda_{\nu}^{IJ}$ are real and diagonal, $M_{N}^{2}={\rm diag}(m^2_{N^k})$ and $\Lambda_{\nu}^{IJ}={\rm diag}(A_{\snu_{k}})$
($k=e,\mu,\tau$ are the flavor index as in Sec.~\ref{sec:param}). The sneutrino mass potential is defined in the sneutrino interaction basis, $\Phi^\dag=(\tilde{\nu}_L^\ast,\, \tilde N^\ast)$, as
\begin{eqnarray}
V_{\rm mass}^k =\frac{1}{2}\, \Phi^{\dag}_{LR}\, \mathcal{M}^2_{LR}\, \Phi_{LR}\,,
\end{eqnarray}
with the squared--mass matrix $\mathcal{M}^2_{LR}$ 
\begin{eqnarray}
&&\mathcal{M}^2_{LR}  = \\ \nonumber
& &\left(  \begin{array}{cc}
m^2_{L^k} + \frac{1}{2} m_{Z}^{2} \cos(2\beta) + m_D^2  & \; \; \; \frac{v\sin\beta}{\sqrt{2}} A_{\snu_{k}}^2  - \mu \frac{m_D}{\tan\beta} \\
                 \frac{v\sin\beta}{\sqrt{2}} A_{\snu_{k}}^2  - \mu \frac{m_D}{\tan\beta}   & m^2_{N^k} + m_D^2
                 \end{array}\right)  \,.
  \label{eq:masslr}
\end{eqnarray}

Here, $m^2_{L^k} \equiv m_L^2$ are the soft mass terms for the three SU(2) leptonic doublets supposed to be common to all flavors, $\tan\beta = v_u/v_d$ and $v^2=v_u^2+v_d^2=(246\,  {\rm GeV})^2$, where $v_{u,d}$ are the Higgs vacuum expectation values (vevs).

The off-diagonal term determines the mixing between LH and RH fields. The Dirac neutrino mass $m_D$ can be safely neglected as it is very small. All the mixing is hence provided by the trilinear term. If it is aligned to the neutrino Yukawa ($A_{\snu_{k}} = \eta Y_\nu$), this term is negligible compared to the diagonal entries and no efficient mixing is provided. However, $A_{\snu_{k}}$ can in general be a free parameter~\cite{Borzumati:2000mc,Arkani-Hamed:2000bq} and can induce a sizable mixing among the sneutrino interaction eigenstates. The sneutrino mass eigenstates are:
\begin{eqnarray}\label{lr_eigenstates}
\left(\begin{array}{c} \snu_{k_1} \\ \snu_{k_2} \end{array}\right) = 
\left(\begin{array}{cc} -\sin\theta^k_{\snu} &   \cos\theta^k_{\snu} \\ 
                                        \cos\theta^k_{\snu} & \sin\theta^k_{\snu} \end{array}\right) 
\left(\begin{array}{c} \snu_L^k \\ \tilde{N}^k \end{array}\right) \,.
\end{eqnarray}
Notice that scalar trilinear term for the sneutrino is connected to the mixing angle as:
\begin{equation} 
 \sin 2\theta_{\snu_k} = \sqrt{2} \frac{A_{\snu_{k}}\, v \sin \beta}{(m^2_{\snu_{k 2}}-m^2_{\snu_{k 1}})}\, .
\end{equation}
The sneutrino coupling to the $Z$ boson, which does not interact with SU(2)$_L$ singlets, is largely reduced if the mixing between LH and RH is sizable. This is extremely relevant to have a good sneutrino DM candidate compatible with the LUX exclusion bound, see {\it e.g.}, Refs.~\cite{Arkani-Hamed:2000bq,Grossman:1997is,Arina:2007tm,Belanger:2010cd}. The limits from direct detection experiments thus restrict the mixing angle for sneutrino DM.

The renormalization group equations are modified by the new singlet superfields $\hat{N}$ as described in~\cite{Belanger:2010cd,Belanger:2011ny}. In particular, the new trilinear term induces an additional running to the LH part, while this is the only correction for the RH soft mass. Assuming common scalar masses and trilinear couplings for all flavors and  neglecting all lepton Yukawas but $Y_\tau$ in the RGEs, the sneutrino tau, $\mlsp$, ends up to be the lightest one among the three sneutrino flavors and hence the LSP, while muon and electron sneutrino remain mass degenerate. 
 
It should be noted that the mass splitting between $\mlsp$ and $\snu_{e_1,\mu_1}$ is frequently smaller than 5 GeV, resulting in practically degenerate sneutrino spectrum at the LHC. However on cosmological time scale the heavier sneutrinos will eventually decay into $\mlsp$, hence they do not play a role at the present epoch for annihilation in astrophysical objects. The rest of the SUSY particles are unaffected by the presence of the RH scalar fields and their behavior is the same as in the MSSM.

\begin{figure}[t]
\centering
\includegraphics[width=0.4\columnwidth]{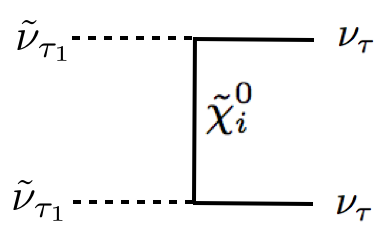}
\caption{Feynman diagram leading to the monochromatic neutrino line: a pair of sneutrinos is annihilating into $\nu_\tau$ via $t$-channel neutralino ($i=1,...,4$) exchange (there is an analogous contribution from pair of anti-sneutrinos annihilating into anti-neutrinos). 
\label{fig:diag}}
\end{figure}

\subsection{Simplified model for the neutrino monochromatic line}\label{sec:paramest}
For WIMP annihilation at the present epoch there are two possibilities for  producing a neutrino flux. The neutrino flux is usually considered to be given by secondary neutrinos resulting from the annihilation of sneutrinos into SM particles ({\it i.e.} $\snu_{\tau_1} \snu^{\ast}_{\tau_1} \to W^+W^-,  ZZ, hh, t\bar{t}$ etc.), which subsequently fragment and decay, leading to neutrinos.  A second possibility is direct tree level (whenever possible) annihilation into monochromatic neutrinos, which will be the main part of this paper.  Figure~\ref{fig:diag} illustrates the tree-level Feynman diagram relevant for sneutrino pair annihilation into a monoenergetic neutrino line. It only involves sneutrino and neutralino parameters, hence to study the behavior of the line with respect to the theoretical parameters of the model, we take a simplified model approach involving sneutrino LSP and the lightest neutralino ($i=1$) for simplicity (this will turn out to be a good assumption when studying the realistic MSSM+RN). We neglect the DM constraints for now. The features of neutrino lines as well continuum for a viable sneutrino DM will be discussed in Sec.~\ref{sec:res}.

On top of the SM particle content, the simplified model we consider is constituted by the sneutrino LSP $\snu_{\tau_1} $, its mixing angle $\theta_{\snu_{\tau}}$, the lightest neutralino $\tilde{\chi}_1^0$ together with its composition: the Bino and Wino fraction ($N_{11}$ and $N_{12}$ respectively) and the Higgsino fraction ($N_{13}, N_{14}$). In total there are two free parameters in the sneutrino sector, the LSP mass $\mlsp$ and $ \sin\theta_{\snu_{\tau}} $, plus 4 free parameters from the electrowikino sector.
\begin{figure*}[t!]
\begin{minipage}[t]{0.5\textwidth}
\centering
\includegraphics[width=1.\columnwidth]{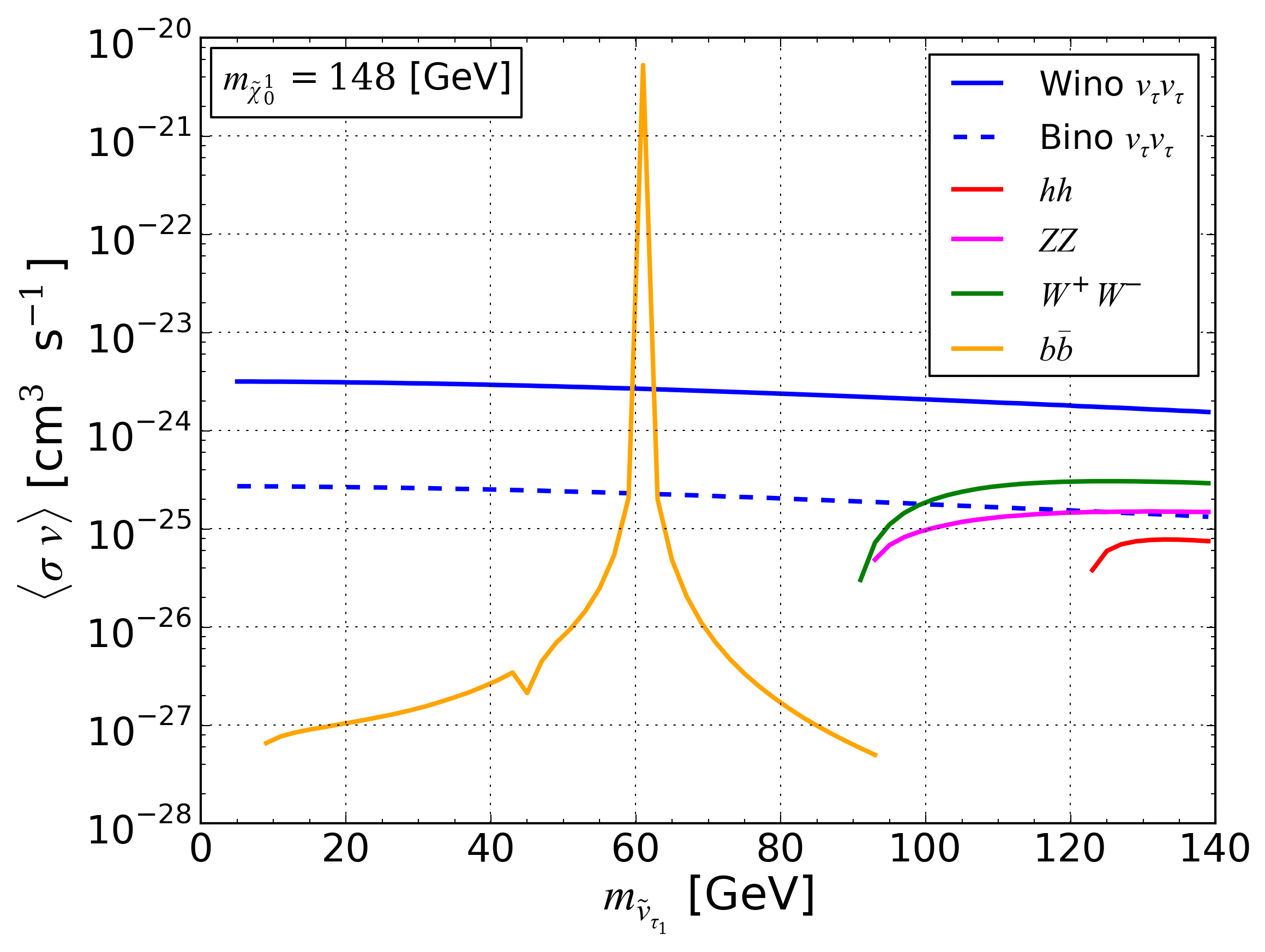}
\end{minipage}
\hspace*{-0.2cm}
\begin{minipage}[t]{0.5\textwidth}
\centering
\includegraphics[width=0.945\columnwidth]{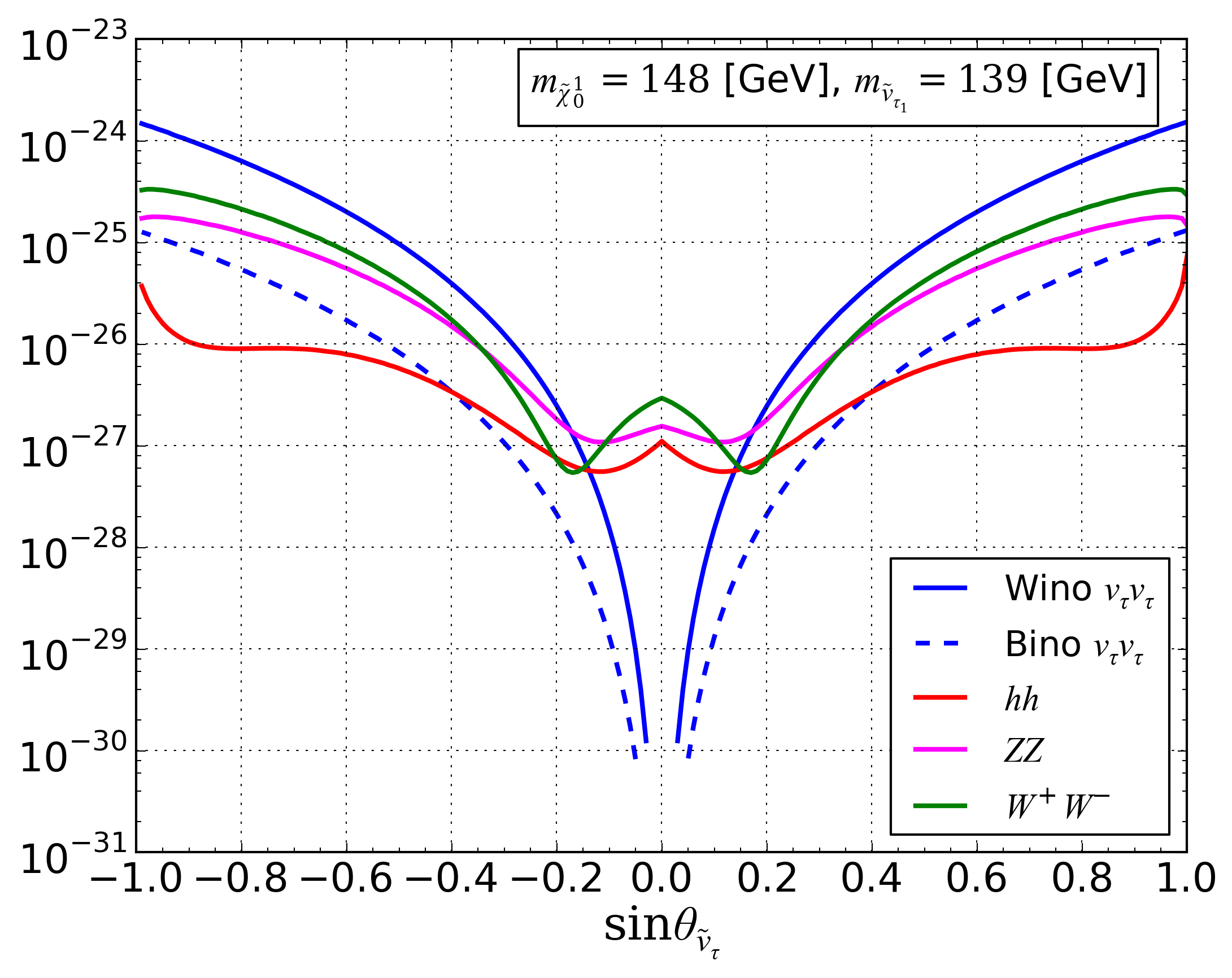}
\end{minipage}
\caption{{\it Left:} The thermal averaged cross-section as a function of the sneutrino mass for each possible sneutrino annihilation channel. The sneutrino is assumed to be pure LH. For fixed mass of neutralino and mixing angles in the neutralino and sneutrino sectors, we vary the mass of the sneutrino. {\it Right:} The thermal averaged cross-section as a function of the sneutrino mixing angle; the trilinear coupling is fixed at 60 GeV and the SUSY masses are fixed as labelled. For fixed neutralino and sneutrino masses and  mixing in the neutralino sector, we vary the mixing angle of the sneutrino. 
\label{fig:paramest2}}
\end{figure*}

The process responsible for the monochromatic neutrino line is $\snu_{\tau_1} \snu_{\tau_1} \to \nu_{\tau} \nu_{\tau}$ via $t$-channel exchange of a neutralino (see Fig.~\ref{fig:diag}). The dominant term for the thermal averaged cross-section is s-wave and given by, see also~\cite{Lindner:2010rr}:
\begin{equation}\label{eq:sigmavsnu}
\sigmav_{\nu_\tau \nu_\tau} = \frac{C_{P_L}^2 + C_{P_R}^2}{8 \pi} \frac{m_{\tilde{\chi}_1^0}^2}{(m_{\tilde{\chi}_1^0}^2 + m_{\snu_{1}}^2)^2} \, ,
\end{equation}
with
\begin{eqnarray}\label{eq:coups}
C_{P_R}  &  = &    - \frac{1}{2} \sin\theta_{\snu_{\tau}}  Y_{\nu}  N_{14}  \,,\nonumber\\
 C_{P_L}  &   = &   - \frac{1}{2} \cos\theta_{\snu_{\tau}}  Y_{\nu}  N_{14} + 
  \frac{e\, \cos\theta_W  N_{11}   \sin\theta_{\snu_{\tau}} }{2 \sqrt{2} (1 - \sin^2\theta_W)}  +  \\
 &&  \frac{e\,  \sin\theta_W  N_{12}   \sin\theta_{\snu_{\tau}}}{2 \sqrt{2} (1 - \sin^2\theta_W) }  -
 \frac{e\,   N_{12}   \sin\theta_{\snu_{\tau}} }{2 \sqrt{2} \sin\theta_W (1 - \sin^2\theta_W) }\,. \nonumber
\end{eqnarray}

$Y_{\nu}$ is the neutrino Yukawa coupling, $e$ the electric charge and $\sin\theta_W$ the Weinberg angle. Since the exchanged particle is Majorana the annihilation process involves as well  two anti-sneutrinos, which will give rise to two monochromatic anti-neutrinos. It is worth noting that there is a velocity dependent subdominant term in $\sigmav$, which can result in enhancement of the neutrino line in case of rotating BHs as mentioned earlier. We neglect this possibility for the current work.
\begin{figure*}[t!]
\begin{minipage}[t]{0.32\textwidth}
\centering
\includegraphics[width=1.\columnwidth]{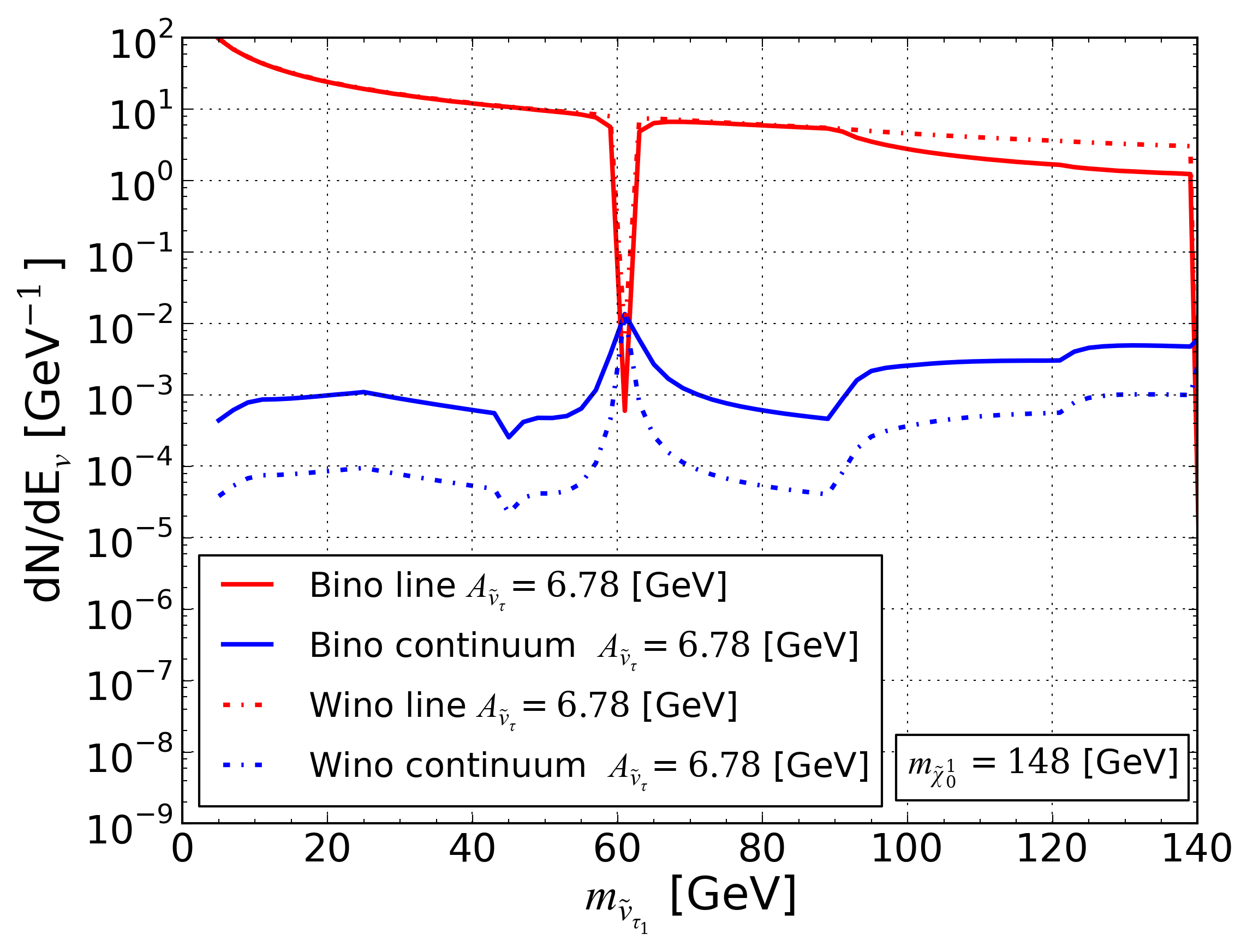}
\end{minipage}
\begin{minipage}[t]{0.32\textwidth}
\centering
\includegraphics[width=0.945\columnwidth]{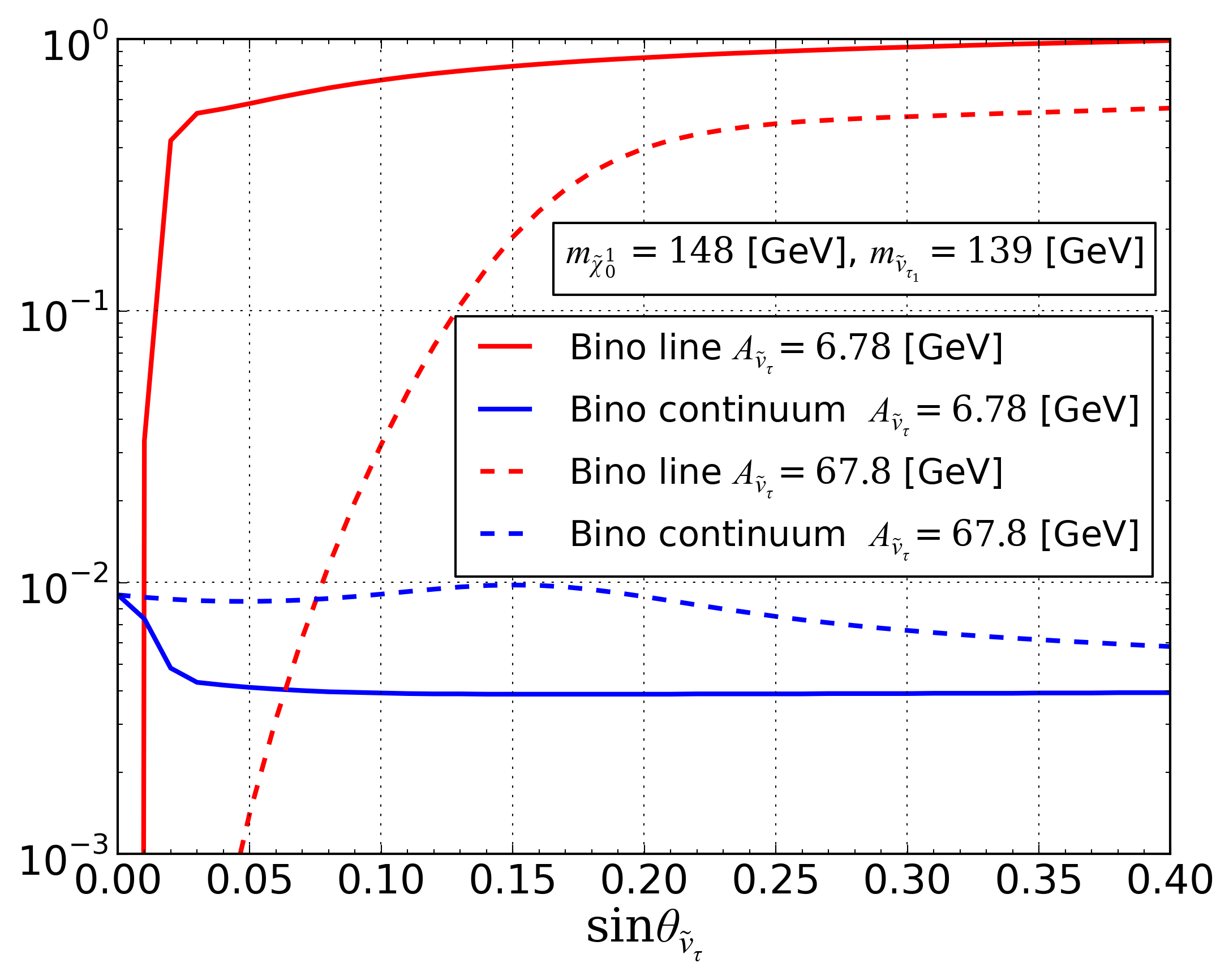}
\end{minipage}
\begin{minipage}[t]{0.32\textwidth}
\centering
\includegraphics[width=0.945\columnwidth]{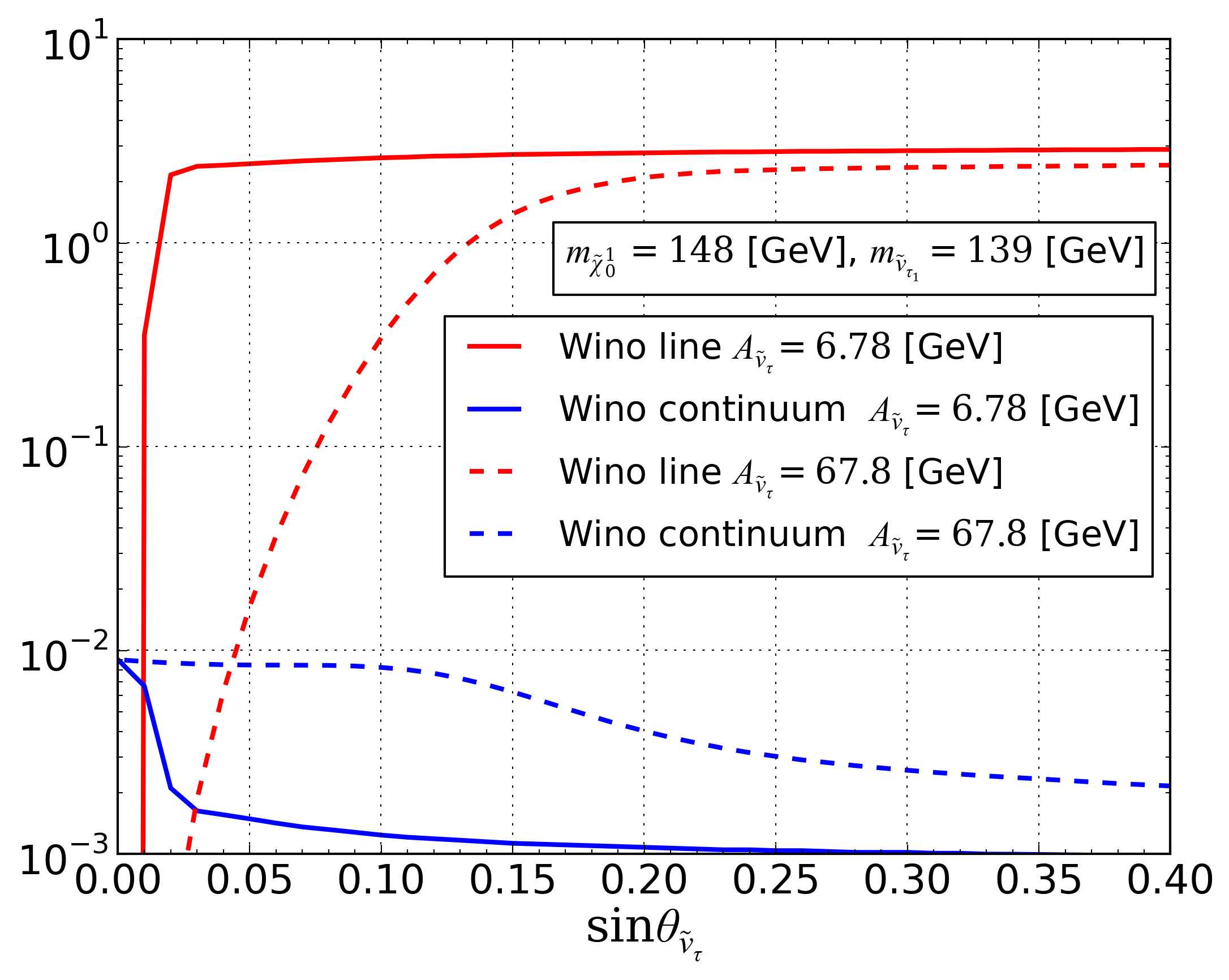}
\end{minipage}
\caption{{\it Left:} Differential neutrino spectrum per annihilation event as a function of purely LH sneutrino mass. The red curves denote the differential neutrino line spectrum, while the blue curves stand for the maximum of the continuum energy spectrum. Both Bino and Wino neutralino cases are considered with solid and dashed lines respectively. {\it Center:} Differential neutrino spectrum per annihilation event as a function of the sneutrino mixing angle. The mass spectrum is fixed as labelled and the neutralino is pure Bino. Two different cases for the trilinear scalar coupling are considered with solid and dashed lines. {\it Right:} Same as the central panel for Wino like neutralino. In all the three panels we vary the parameter on x-axis while keeping the rest of the parameters fixed. The mixing angle is varied independent of the trilinear coupling. 
\label{fig:paramest1}}
\end{figure*}

From Eq.~\ref{eq:sigmavsnu}, we notice that $\sigmav$ is directly proportional to the mass difference between sneutrino and neutralino. Indeed for $\mneut >> \mlsp$ the thermally-averaged cross-section simplifies to $\sigmav_{\nu_\tau \nu_\tau} \propto 1/\mneut^2$. From Eq.~\ref{eq:coups}, it is clear that the Higgsino component does not contribute to this process as it suppressed by the neutrino Yukawa coupling. This also implies $C_{P_R} \sim 0$.\footnote{$C_{P_R} \sim 0$ is an artifact of this particular model of sneutrino LSP, with the Dirac mass term only. In the case of Majorana masses, one could get a non-trivial contribution from $C_{P_R}$.} In order to illustrate the dependence of annihilation cross-section and the differential spectrum of the neutrino line, we vary the physical masses of the neutralino and sneutrino as well as their mixing angles, independently of each other. The dependence of  Eq.~\ref{eq:coups} on the Bino and Wino fractions of $\tilde{\chi}^0_1$ is displayed in Fig.~\ref{fig:paramest2}, together with the other annihilation channels of the sneutrino. The neutrino line final state is the only one that depends on other SUSY particles while all the other channels involve couplings and masses of SM particles or that of additional Higgs bosons, once the sneutrino mixing angle is fixed. In Fig.~\ref{fig:paramest2}, left panel, we choose a purely LH sneutrino ({\it i.e.} $|\sin\theta_{\snu_{\tau}}| = 1$) and plot $\sigmav$ for $\mlsp < \mneut$. As soon as the sneutrino mass approaches the neutralino mass $\sigmav$ decreases, while the overall normalization is dictated by the neutralino composition. Clearly the Wino case is enhanced  with respect to the pure Bino case, as it receives contributions from both the SU(2)$_L$ and U(1) couplings. As soon as kinematically allowed, the annihilation channels into $W^-W^+$ and $ZZ$ open up,  mediated by $s$-channel exchange of a $Z$ boson for LH sneutrinos. The $h h$ final state comes from the scalar quartic vertex with the sneutrinos, while for the whole mass range shown the annihilation channel into fermions is always open (we only show the $b\bar{b}$ contribution for simplicity). The $b\bar{b}$ final state receives large contributions from the Higgs pole and a small enhancement at the $Z$ pole. We do not plot the continuum for Wino case as it is independent of the neutralino nature. The right panel exhibits the dependence of $\sigmav$ on the sneutrino mixing angle. $C_{P_L}$ is directly proportional to $\sin\theta_{\snu_{\tau}}$, hence for purely RH sneutrino the annihilation into monochromatic lines is not possible. The relative weight between different annihilation channels strongly depends on the mixing angle, the contribution from Higgs $s$-channel becomes important relative to the $Z$ boson $s$-channel exchange as far as $\sin\theta_{\snu_{\tau}}$ decreases. The annihilation channels involving the Higgs boson are also proportional to the value of the scalar trilinear coupling, which appear in the Higgs coupling, see Eq.~\ref{lrsoftpot}.  Those considerations find further support from Fig.~\ref{fig:paramest1}, where the differential neutrino spectrum per annihilation event is shown. The first panel illustrates the dependence of ${\rm d}N/{\rm d}E_{\nu_{\rm line}}$ (red) and of the maximum of the $\nu_{\tau}$ continuum (blue) as a function of the sneutrino mass, assuming again a LH LSP. The line is enhanced over the continuum over all of the mass range, except at the Higgs pole, when sneutrino annihilation into $b\bar{b}$ dominates the thermally averaged cross-section. The suppression of the differential neutrino line spectrum is negligible at the Z pole. The continuum seems to depend on the neutralino nature, however this is an artifact of the overlap of the Bino and Wino cases for the line. Since the Wino contribution is larger, then the neutrino continuum differential spectrum decreases. ${\rm d}N/{\rm d}E_{\nu_{\rm line}}$ for both Bino and Wino neutralino overlap until the opening of the $W^+W^-$ final state, after which the Bino case has a smaller contribution. The central and right panels show the dependence of ${\rm d}N/{\rm d}E_{\nu}$ on the sneutrino mixing angle. The line intensity (red lines) drops abruptly for $\sin\theta_{\snu_{\tau}} \to 0$, while it reaches a plateau value for sizable mixings and is larger than the continuum.  We show the behavior for two different values of the scalar trilinear term: as expected the larger $A_{\snu_{\tau}}$ the smaller the contribution of the line with respect to the continuum. This holds for both the Bino (central panel) and Wino (right panel) cases. 

In summary, the conditions to have a sizable contribution from the process $\snu_{\tau_1} \snu_{\tau_1} \to \nu_\tau \nu_\tau$ with respect to $\snu_{\tau_1} \snu_{\tau_1}^\ast$ annihilations are: 1) sizable mixing angle, 2) small $A_{\snu_{\tau}}$, 3) sizable mass gap between LSP and lightest neutralino and 4) relatively light mass spectra.

In the case of the neutralino LSP, the neutrino lines are further suppressed because of the Majorana nature of the initial state. The process is analogous to Fig.~\ref{fig:diag} with the neutralinos exchanging a sneutrino in $t$-channel, with an expression equivalent to Eq.~\ref{eq:sigmavsnu} up to  smaller normalization factors. Notice however that in this case the sneutrino can be purely LH and the line is not bound to be a monochromatic $\nu_{\tau}$ line but could involve all three neutrino flavors. Because of the smallness of the line in the case of the neutralino LSP, we do not investigate any further the possibility (see Ref.~\cite{Lindner:2010rr} for details).
%

\section{Detection of monochromatic neutrino lines}\label{sec:set}

The neutrino spectra produced by DM annihilations at the source (SMBHs or dSphs) position are shown in Fig.~\ref{fig:examples}, left panel. The neutrino with $\tau$ flavor has a sharp line at the DM mass and a negligible continuum spectrum (red), contrary to the case of $\nu_e$ and $\nu_\mu$ (blue and green). The $\nu_e$ and $\nu_\mu$ only contribute to secondary neutrino flux and do not produce line. These spectra however are modified by neutrino flavor oscillations as follows.

\subsection{Neutrino flavor oscillations in short}\label{sec:osc}
\begin{figure*}[t]
\centering
\includegraphics[width=0.7\textwidth]{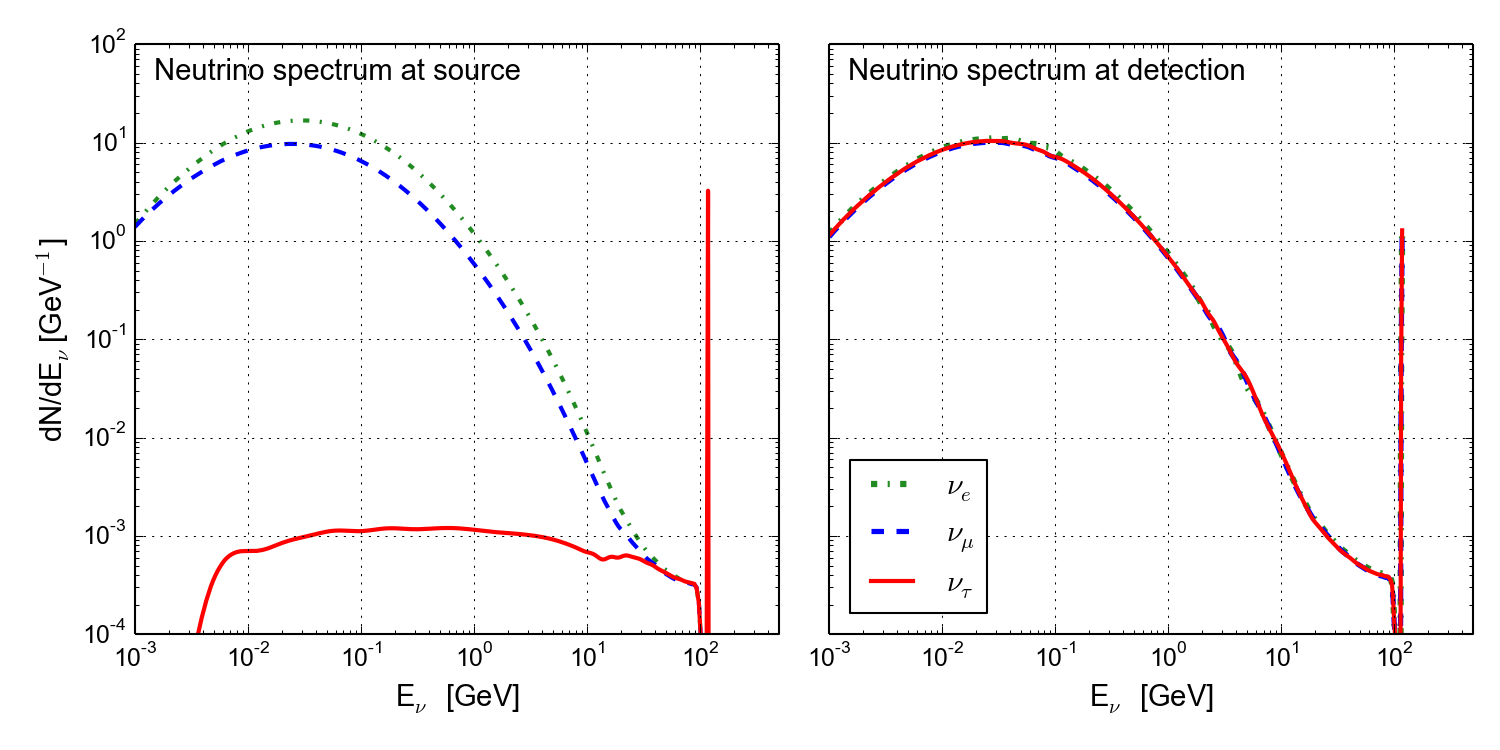}
\caption{{\it Left:} Differential neutrino spectra per annihilation event as a function of the neutrino energy, at the source. {\it Right:} Same as left but at detection, after including vacuum and matter oscillation thought the Earth; normal hierarchy and zenith angle  $\cos\delta= -1$ are assumed. A very similar spectrum arises for the antineutrinos. The parameters for this point are $\mlsp = 118$ GeV, $\mneut =127$ GeV, $N_{12} = 0.996$, $N_{11} = 0.001$, $\sigmav = 2.15 \times 10^{-31} \rm cm^3 s^{-1}$, $\bra_{\rm line} = 0.38$, $A_{\snu_{\tau}} = 6.78$ GeV, $\sin\theta _{\snu_{\tau}}= -0.024$. 
\label{fig:examples}}
\end{figure*}

The propagation of neutrinos produced by sneutrino annihilations is affected by flavor oscillations, first in vacuum, when they travel from the dSphs or the SMBHs to the Earth, and eventually by oscillations in matter when crossing the Earth to reach the neutrino telescope. 

We follow~\cite{PDG} for the implementation of flavor oscillations in vacuum in the long baseline limit, as the flavor oscillation length is negligible compared to the distance of the astrophysical source. As for flavor oscillations in matter, we use the results from~\cite{Baratella:2013fya} together with their choice of oscillation parameters for coherence.\footnote{These are up-to-date values and take into account a 
non-zero mixing angle between the neutrino mass eigenstates 1 and 3, {\it i.e.} $\theta_{13} = 0.15$.}

The neutrino spectra we use in the rest of the analysis assume normal hierarchy for neutrino masses and a maximal crossing of the Earth (zenith angle $\cos\delta=-1$). This last condition is more suitable to produce up-going events into the neutrino detectors. An example of the neutrino energy distribution after oscillations is shown in the right panel of Fig.~\ref{fig:examples}: with respect to the injection neutrino spectrum (left), the overall effect of flavor oscillations is the democratic redistribution of the line in all three neutrino flavors, namely ${\rm d}N/{\rm d}E_{\nu_{\rm line}}^k \sim 1/3\,   {\rm d}N/{\rm d}E_{\nu_{\rm line}}^{\rm source}$ ($k=e,\mu,\tau$).

We do not provide further details on this part of the discussion, as different assumptions on the neutrino mass hierarchy do not have significant impact on the results of our analysis. The results for anti-neutrinos, are similar to those for  $\nu$s.

\subsection{Future neutrino telescope set-up}\label{sec:exp}

Small values for the mixing angle are closer to `realistic' sneutrino DM scenarios. It is indeed well known that pure LH sneutrinos as DM are excluded by direct detection constraints, and only very small mixing angle are allowed by LUX, as we will see in the next section. Already in the simplified model framework we can see from Fig.~\ref{fig:paramest2} that small $\sin\theta_{\snu_{\tau}}$ imply values of  $\sigmav$ below $10^{-27} \rm cm^3 s^{-1}$. These values of $\sigmav$ are orders of magnitudes below current constraints arising from DM searches at the galactic center by neutrino telescopes. IceCube (IC) detector holds the most stringent upper bound~\cite{Aartsen:2015xej} on the thermally averaged cross-section for DM masses between $10^2 - 10^3$ GeV, to be less than  $\sigmav \sim 5 \times 10^{-22} \rm cm^3 s^{-1}$. 

In this section we propose a search strategy for monochromatic neutrinos from DM annihilations and provide a guideline for the ideal set-up of future neutrino detectors. We focus on the energy range between 100 GeV to TeV DM mass scales,  one of the most common ranges predicted by scenarios beyond the Standard Model, and assume that the values of $\sigmav$ are below the reach of present DM detectors. Such small values of $\sigmav$ are characteristic of the sneutrino scenario we use as a proxy for the monochromatic line search, but can be realised in several other WIMP scenarios as well.

The IC neutrino telescope consists of 5600 digital  optical modules (DOMs)  arranged  vertically  along 86  strings  embedded  in  a  km$^3$   of  extremely transparent   ice  below  the  South  Pole. The DeepCore array constitutes the central part of IC and consists of six strings arranged with closer horizontal spacing and instrumented with DOMs of higher sensitivity and closer vertical spacing along the string.  DeepCore is able to reduce the energy threshold of IC from 100 GeV down to 10 GeV, however the energy resolution for these energies, without directional information like in the case of searches for DM from the Sun, is estimated to be not more than 50\%. In general, it should be noted that neutrino detectors such as IC and Antares~\cite{Adrian-Martinez:2015wey}, the largest neutrino detector in water in the Northern Sky, have indeed been designed and optimised for ultra high energy neutrinos ($E_{\nu} > $ few TeVs) and hence are not sensitive to the range of masses considered in this paper.  

The detection of neutrinos occurs via the observation of the Cherenkov radiation emitted by the particles produced in the interactions between the neutrinos and the matter in and around the telescope. Particularly interesting are muons created by the charged current interactions of $\nu_\mu$, as they create long and relatively detectable tracks compared to other leptons.  An even more clean signature is that of  up-going muons, which provide a powerful discriminant to distinguish galactic and extra galactic neutrinos from atmospheric neutrinos. By considering these up-going events, current neutrino detectors have a rather high sensitivity to point sources in the opposite hemisphere emitting ultrahigh energy neutrinos. For instance, IC40\footnote{IC40 means that the analysis has been done with data taken by IceCube with 40 strings.}~\cite{Aartsen:2013uuv} has a sensitivity down to $E\,  \Phi_\nu \sim 2 \times 10^{-9} \rm GeV cm^{-2} s^{-1}$ to sources in the Northern sky, while Antares \cite{Kouchner:2015mra} can probe down to $E\,  \Phi_\nu \sim 1.5 \times 10^{-8} \rm GeV cm^{-2} s^{-1}$ sources in the Southern sky.  

We first assume that a future generation of neutrino telescopes can achieve similar sensitivities as IC40 and Antares to point sources, for neutrinos in the energy range considered in the paper. In other words neutrino telescopes with the same effective area as for  $>$ TeV  for 100 GeV $-$ TeV DM particles would better perform for DM searches.\footnote{The effective area is by definition the area for which detection efficiency of a neutrino is 100\%.} This roughly  implies detectors of the current size, however with an increased string configuration and granularity of the detector, similarly to DeepCore, to augment the energy resolution. Once these requirements are met, the most promising targets for WIMP searches  are to look for DM in overdense regions, such as dSPhs and DM spikes from central BHs in galaxies. 

In our analysis, we provide the expected flux of monochromatic muon neutrinos, to rely on up-going muon track detection, from DM annihilations  (notice that $\nu_e$ and $\nu_\tau$ will provide a similar flux of up-going events giving rise to shower events instead of muons tracks) and compare it with the neutrino telescope sensitivity. The IC40 ideal detector can thus probe M87, NGC1277 and the dSphs in the Northern sky, while the Antares-like detectors would probe CenA and the dSphs in the Southern sky. Very  optimistically, we give the expected sensitivity of the upgraded version of IC, a 10 km$^3$ detector called IceCube-Gen2 (IC-Gen2), after 5 years of running~\cite{Aartsen:2014njl}, and for the water km$^3$ detector KM3Net~\cite{Sapienza:2013wla} in the Northern hemisphere as upgrade of Antares. We assume that the energy resolution is about $30\%$, as it is expected for $\mathcal{O}(1)$ TeV neutrinos, and an angular resolution of $1^{\circ}$.  

Although our estimates of the telescope sensitivities are optimistic, our aim here is to analyse the potential of future  neutrino telescopes for detecting neutrino lines. We do not embark on a detailed study of the detection prospects in this work, neither do we attempt to give any details of the possible telescope design required in order detect such a signal.

%
%
%
%

\section{Sneutrino DM from MSSM+RN: realizations for the neutrino monochromatic lines}\label{sec:res}
In this section, we present the prospects for detection of mono-chromatic neutrino lines in the full MSSM+RN model. 

The input parameters for the MSSM+RN are non-universal gaugino masses, the two Higgs doublet masses, $\mu$ and $B$ are respectively the mass term for the Higgs fields in the superpotential and soft breaking potential, Eqs.~\ref{lrsuppot} and~\ref{lrsoftpot}, the trilinear couplings for charged sleptons, sneutrinos and squarks and the soft scalar masses for LH sleptons and squarks, for RH charged sleptons, RH sneutrinos and squarks. All MSSM+RN input parameters are defined at the Grand Unification scale $\sim 10^{16}$ GeV. These 13 free parameters are efficiently sampled with the nested sampling package \texttt{MultiNest\_v3.2}~\cite{Feroz:2007kg,Feroz:2008xx}, based on Bayes theorem.  As a result, the sampled points are distributed accordingly to the posterior probability density function and satisfy the DM constraints: relic density constraints from Planck measurements~\cite{Planck:2015xua} and the elastic spin independent (SI) scattering cross-section $\sigma_{\rm Xe}^{\rm SI}$ is compatible with the 90\% CL of the LUX exclusion bound. All details about the sampling procedure, the constraints and measurements implemented in the likelihood function, and the LHC phenomenology are provided in~\cite{Arina:2015uea}. At practical level the relic density and direct detection are computed with \texttt{micrOMEGAS\_3.6}~\cite{Belanger:2013oya}, while we use \texttt{micrOMEGAS\_4.1}~\cite{Belanger:2014vza} for the neutrino spectrum and indirect detection annihilation cross-section. 
\begin{figure}[t]
\centering
\includegraphics[width=1.\columnwidth,trim=0mm 7mm 0mm 17mm, clip]{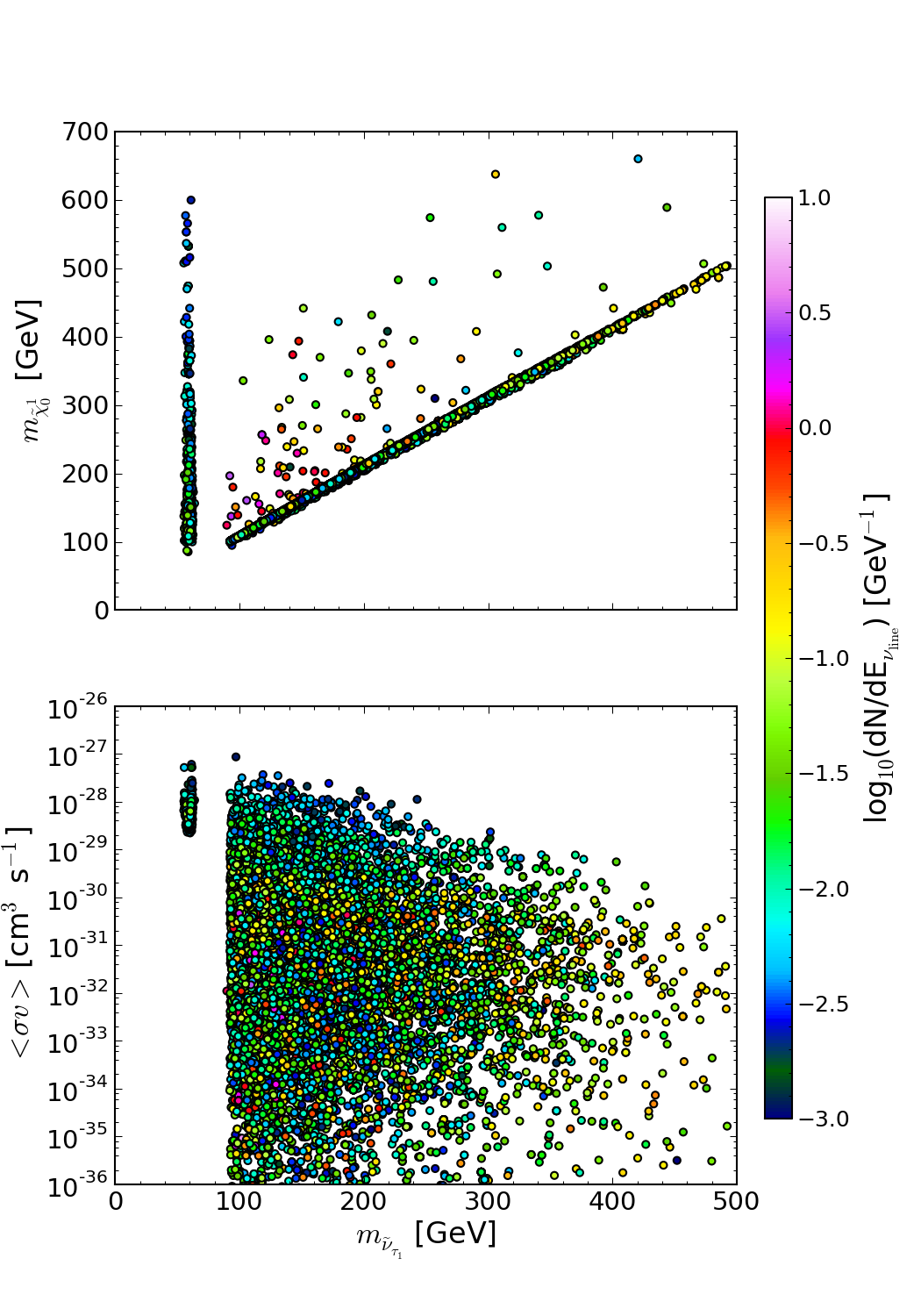}
\caption{{\it Top:} The differential energy spectrum per annihilation event of $\nu_\tau$ at the source position is shown as a function of both the sneutrino mass and the neutralino mass. {\it Bottom:} Same as top as a function of sneutrino mass and annihilation cross-section. The color code for ${\rm d}N/{\rm d}E_{\nu_{\rm line}}$ is given by the color bar on the right hand side.
\label{fig:param2}}
\end{figure}
\begin{figure}[t]
\centering
\includegraphics[width=1\columnwidth,trim=2mm 2mm 1mm 2mm, clip]{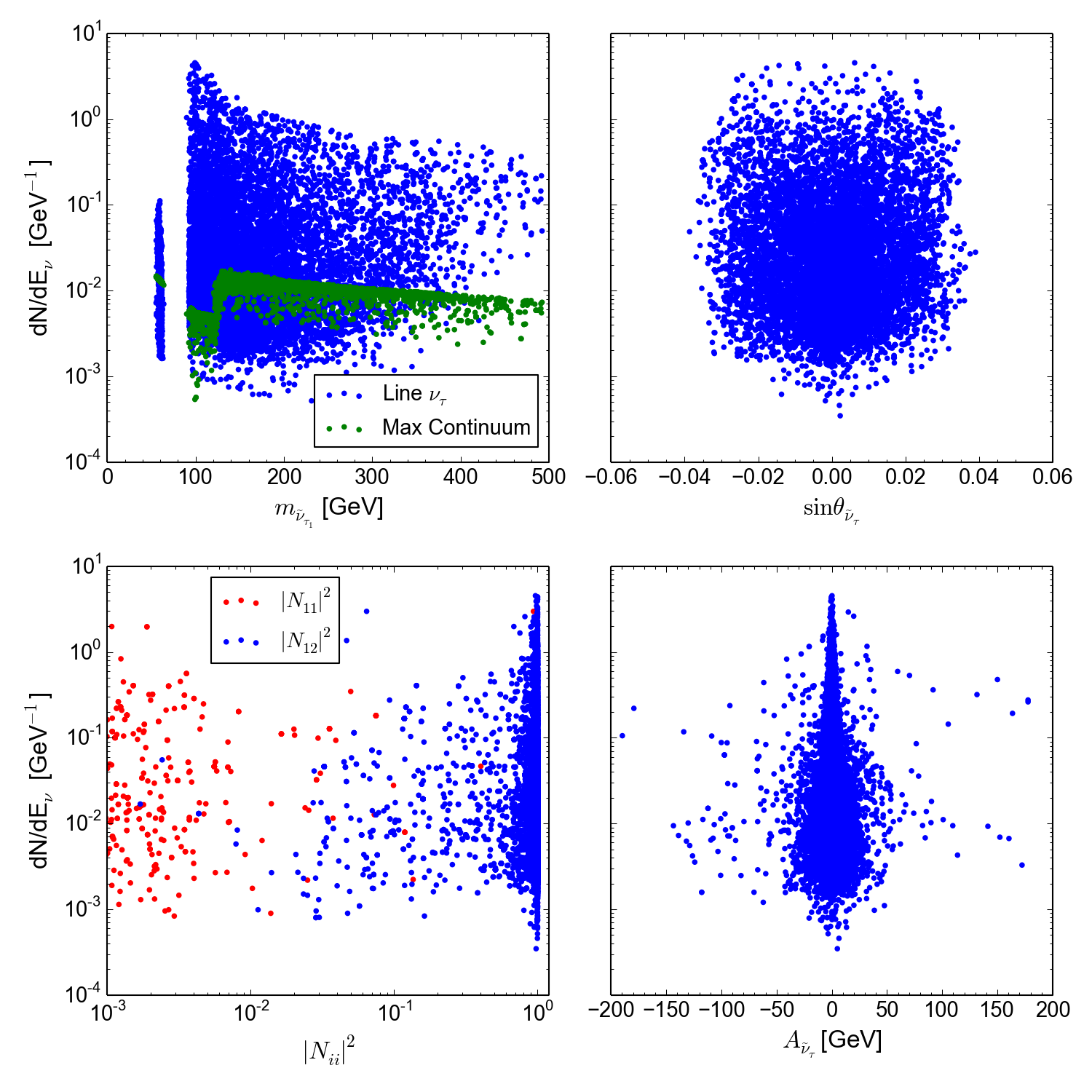}
\caption{{\it Upper left:} Differential energy spectrum for the $\nu_\tau$ line (blue) and for the continuum (green) are shown as a function of the sneutrino mass. {\it Right:} Same as left as a function of the sneutrino mixing angle, for the $\nu_\tau$ line spectrum only. {\it Lower left:} Differential $\nu_\tau$ line spectrum as a function of  the Bino (red) and Wino (blue) fractions of the lightest neutralino. {\it Lower right:} Same as top right as a function of the scalar trilinear coupling. In all panels the differential spectrum is given at source.\label{fig:param1}}
\end{figure}

In this section we study the sampled points which have small $A_{\snu_{\tau}}$ terms and reasonably large sneutrino mixing angles, both necessary conditions to have neutrino lines, as discussed in Sec.~\ref{sec:paramest}.\footnote{This sample corresponds to logarithmic prior probability density functions on all parameters.} Of course, all these points satisfy the constraints listed in~\cite{Arina:2015uea}.

The differential spectrum for $\nu_{\tau}$ per annihilation event is shown in Fig.~\ref{fig:param2} in the sneutrino-neutralino mass plane (top panel) and as a function of the sneutrino mass and 
thermally-averaged annihilation cross-section at the present epoch (bottom panel). Considering first the bottom panel it should be noted that the typical value of $\sigmav$ is smaller than the usual value for the standard WIMP relic. Indeed the largest values start around $10^{-27} \rm cm^3 s^{-1}$ and go down even below $10^{-36} \rm cm^3 s^{-1}$.
For the sneutrino mass range covered by the sample, these values of $\sigmav$ would be too low to achieve the correct relic density via its annihilation only ({\it i.e.} for these values the sneutrino would cause overclosure). In the early Universe there is one primary coannihilation process at work: the lightest neutralino coannihilates efficiently with the lightest chargino and achieves the correct relic density ($\tilde{\chi}^0_1\, \tilde{\chi}^\pm_1 \to q \bar{q}', Z W^\pm, ...$ and $\tilde{\chi}^0_1 \tilde{\chi}^0_1 \to W^+W^-$ and $\tilde{\chi}^{+}_1 \tilde{\chi}^{-}_1 \to W^+ W^-, q\bar{q}, Z Z, ...$), which is subsequently transmitted to the sneutrino sector. This argument is further supported by the top panel, where $m_{\tilde{\chi}^0_1}$ is shown as a function of the sneutrino mass. The sneutrino and lightest neutralino are almost degenerate, except at the Higgs pole and for few other points. The proximity in mass between the DM and $\tilde{\chi}^0_1$ tends to cause a suppression in the intensity of the monochromatic $\nu_\tau$ line, denoted by the color bar and the third dimension in both panels. Indeed the largest values for ${\rm d}N/{\rm d}E_{\nu_{\rm line}}$ are achieved for sizable mass differences between the lightest neutralino and the LSP (reddish points).  In the few points above 100 GeV where the mass gap is sizable, the sneutrino can thermalize due to coannihilation processes with $\snu_{\mu_1}, \snu_{e_1}$ and/or with the charged sleptons, most usually the $\tilde{\tau}^\pm$. On the contrary, even if at the Higgs pole, the mass gap between sneutrino and neutralino is large, the channels $\snu_{\tau_1} \snu_{\tau_1}^\ast \to h \to b \bar{b}, \tau^+ \tau^-, ...$ dominate and suppress all the other sneutrino interactions. These features were expected from the discussion on the simplified model for the line, Sec.~\ref{sec:paramest}. 
\begin{figure}[h!]
\centering
\includegraphics[width=1.\columnwidth,trim=0mm 7mm 0mm 17mm, clip]{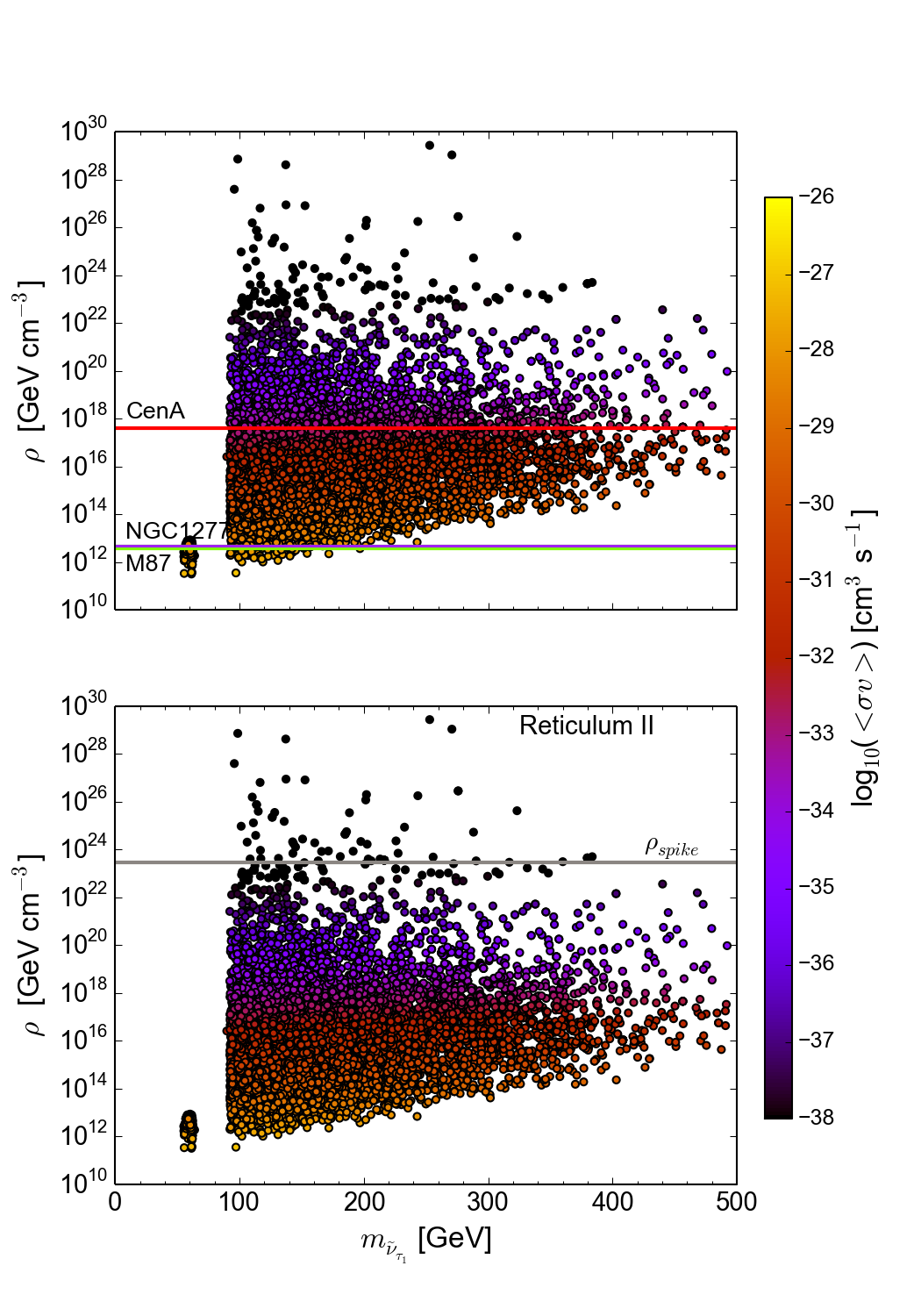}
\caption{{\it Top:} DM density plateau, Eq.~\ref{eq:rhopl}, predicted by the MSSM+RN, as a function of the sneutrino mass and of the annihilation cross-section (third direction denoted by the color code and color bar on the right). Also shown with  horizontal lines are the maximal values of the spike profiles, Eq.~\ref{eq:rhopsike}, for each of the SMBHs used in the analysis, as labelled. {\it Bottom:} Same as top for Reticulum II hosting a $M_{\rm BH} = 10^4 M_{\odot}$ in the inner center. The maximal value of the density spike is indicated by the gray horizontal line.
\label{fig:astrosv}}
\end{figure}
The behavior of the line intensity ${\rm d}N/{\rm d}E_{\nu_{\rm line}}$ is further investigated in Fig.~\ref{fig:param1}, where the dependence on the sneutrino mixing angle and neutralino composition is shown. We show the dependence on the Wino (blue) and Bino (red) fraction of $\tilde{\chi}^0_1$ in the lower left panel. Most of the lightest neutralinos are Wino-like because of the coannihilation requirements in the early universe, {\it i.e.} in order to be within 10\% in mass with the lightest chargino~\cite{Griest:1990kh}. When the neutralino is Bino-like, the relic density is set by cohannihilation with sleptons. In any case, the Higgsino components of the neutralino are always negligible, which positively impacts the process giving rise to the monochromatic neutrino line. The sneutrino mixing angle compatible with the LUX exclusion limit in direct DM searches are given in the top right panel. As one can see these values are pretty small and ranges from practically 0~(i.e. almost pure RH sneutrinos) to the maximal allowed value of $|\sin\theta_{\snu_{\tau}}| \leq 0.04$. As discussed for a couple of benchmarks in Fig.~\ref{fig:paramest2}, ${\rm d}N/{\rm d}E_{\nu_{\rm line}}$ depends on $\sin\theta_{\snu_{\tau}}$ and tends to reach a plateau value. The mixing angle value at which the plateau sets in depends strongly on $A_{\snu_{\tau}}$: larger values of the trilinear scalar coupling tend to delay the onset of the plateau to larger mixing angles. The values of the trilinear coupling are given in the lower right panel: the largest value of ${\rm d}N/{\rm d}E_{\nu_{\rm line}}$ are indeed correlated with the smallest values of $A_{\snu_{\tau}}$ Hence the constraint by the LUX experiment may suppress the overall line intensity because it selects on average mixing angles much smaller than the value required to reach the plateau value for a definite SUSY spectrum configuration. SI elastic scattering of sneutrino off nuclei arises via the exchange of a $Z$ and $h$ boson on $t$-channel. The $Z$ channel is independent of the sneutrino mass and leads to $\sigma^{SI}_{\rm Xe} \sim 10^{-39} \rm cm^2$. Since the LUX bound, for the mass ranges 60 to 500 GeV, ranges from $10^{-45} \rm cm^2$ to $6 \times 10^{-45} \rm cm^2$, the $Z$ exchange should be suppressed by almost 6 order of magnitude, which explains the smallness of $\sin\theta_{\snu_{\tau}}$, while the Higgs exchange is controlled by $A_{\snu_{\tau}}$.

The other important contributions to the neutrino final state are the secondary neutrinos, leading a continuum in the flux. In the first panel of Fig.~\ref{fig:param1} we illustrate the relative weight of the line spectrum with respect to the $\nu_{\tau}$ differential secondary spectrum (green dots) as a function of the sneutrino mass. The origin of the continuum flux is clear: at the Higgs pole it is due to sneutrino annihilations into fermions, while above the $W$, $Z$ and $h$ thresholds it receives contributions from  $W^+W^-$, $ZZ$  and $hh$ final states respectively (the kinematic edges for the latter final state and for the $W^+W^-$ threshold appear particularly clean in the plot). The line signal can be enhanced over the continuum by on average two orders of magnitude, the enhancement being bigger for light sneutrinos and light  but not degenerate $\tilde{\chi}^0_1$ (Fig.~\ref{fig:param2}), as expected. 

The differential spectrum for the monochromatic line discussed until now is given at the source position. As described in Sec.~\ref{sec:osc}, the effect of neutrino flavor oscillations due to propagation from the distant galaxies to Earth is to reduce the line intensity by roughly 1/3 and to distribute it democratically among all flavors. This determines the ${\rm d}N/{\rm d}E_{\nu_{\rm line}}$ at detection.

To estimate the expected neutrino flux, Eq.~\ref{eq:dflux}, it is necessary to estimate $\Phi_{\rm Astro}$, which in turns depends on the particle physics model in case of DM spikes, Eqs.~\ref{eq:rhor} and~\ref{eq:rhopl}, because of the annihilation plateau that can smooth out the DM spike. 
\begin{figure*}[t]
\centering
\includegraphics[width=1.\textwidth,trim=0mm 0mm 5mm 4mm, clip]{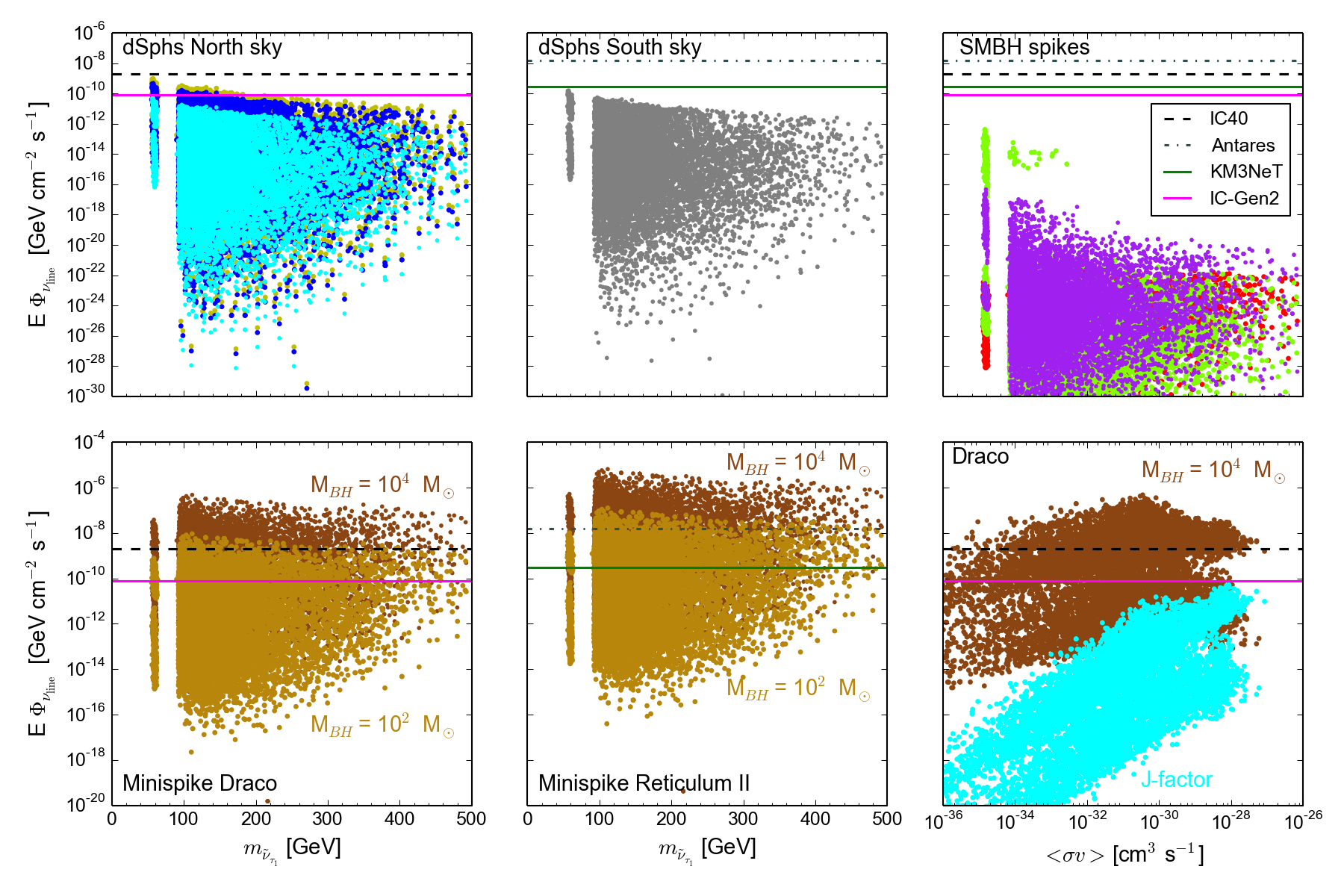}
\caption{Expected flux for the neutrino line as a function of the sneutrino mass, for all astrophysical environments considered in the analysis, as labelled in each panel. The flux is computed for $\nu_\mu$ and is given at the detector position. The color code for the sampled points is as follows. {\it [Upper panels] Left:} Cyan stands for Draco, blue for Ursa Major II and yellow denotes the stack of all dSphs in the Northern Sky. {\it Center:} Gray stands for Reticulum II (equivalent to the stack of dSphs in the Southern Sky). {\it Right:} Red denotes CenA,  while light green and violet stand for  M87 and NGC1277 respectively. {\it [Lower panels] Left and center:} Effect of the minispike for two different BH masses, as labelled. {\it Right:} Expected flux for the neutrino line as a function of the thermal annihilation cross-section in Draco, with(out) minispike in brown (cyan). In all panels we show the sensitivity of current and future neutrino detectors for point sources, assuming that it holds as well for neutrinos with energies $E_{\nu} \sim 100 \to 500$ GeV. 
\label{fig:flux_msnu}}
\end{figure*}
\begin{figure*}[t]
\centering
\includegraphics[width=0.7\textwidth]{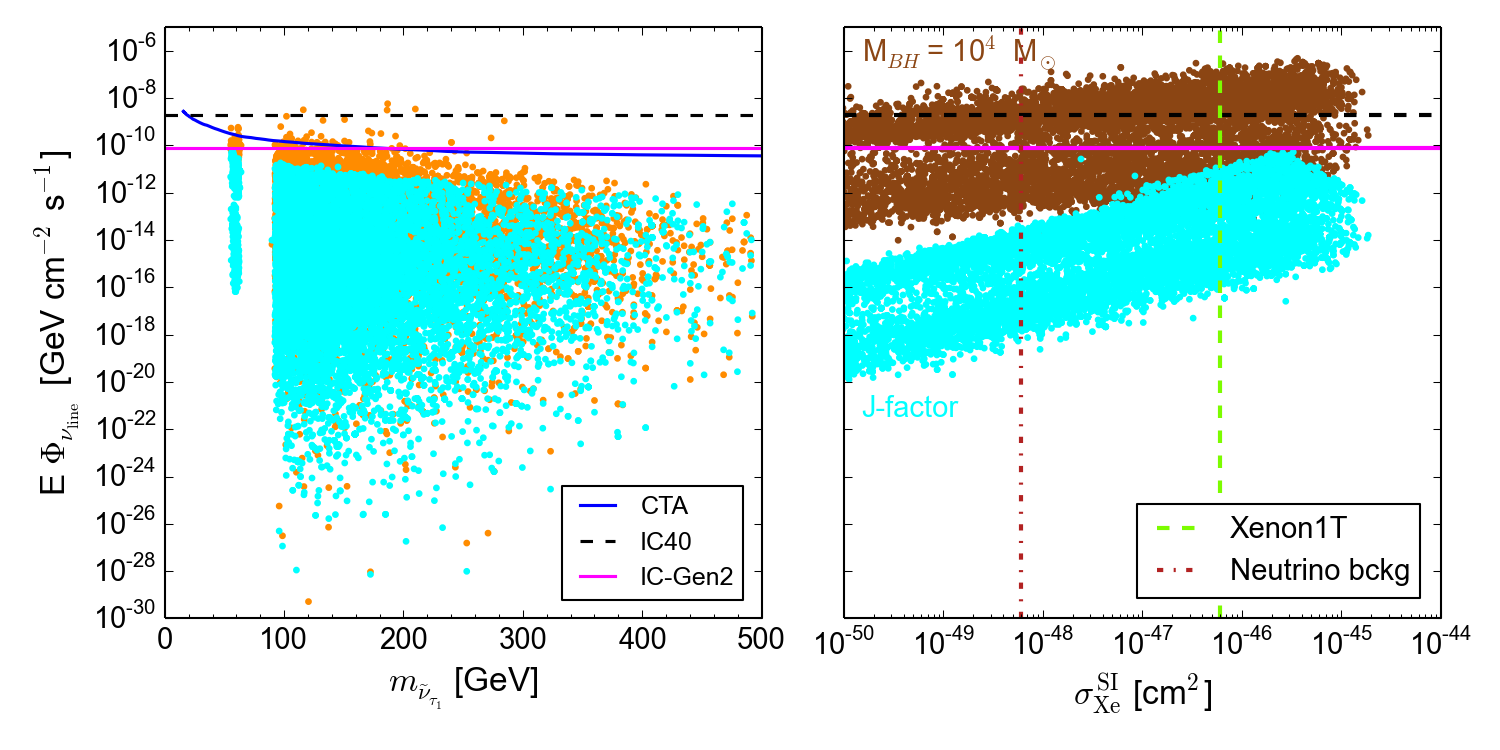}
\caption{{\it Left:} Expected neutrino line flux compared to the expected gamma ray flux (orange) from sneutrino annihilation in Draco, as a function of the sneutrino mass. The point source sensitivity of CTA is shown together with the neutrino telescope sensitivities as labelled. {\it Right:} Predicted SI cross-section versus $\nu_\mu$ line flux. We show the complementarity of indirect and direct DM searches with and without minispike in Draco as labelled. The expected sensitivity of XENON1T and the neutrino background are shown by the vertical lines, as labelled.
\label{fig:flux_msnu2}}
\end{figure*}
Here it should be made clear that the values of $\sigmav$ that enter into computation of the DM density plateau, $\rho_{\rm pl} \propto 1/\sigmav$, are those reported in Figs.~\ref{fig:param2} and~\ref{fig:astrosv}. For sneutrino DM, it is crucial to differentiate between the value of $\sigmav$ in the early universe and at the present time. Those two quantities can differ by several orders of magnitude due to coannihilation effects, as already discussed above. Since after freeze-out, all SUSY particles decay into $\snu_{\tau_1}$, at the time of the SMBH or dSPh formation, the DM is constituted solely of sneutrinos, annihilating with small $\sigmav$. Figure~\ref{fig:astrosv} illustrates the value range for the DM density plateau for the sampled points, as a function of the sneutrino mass and $\sigmav$(denoted by the third direction and color bar on the right). The top panel refers to the SMBH case and shows the maximal value of the spike profile, $\sim 5 \times 10^{12}\rm GeV\,  cm^{-3}$, for M87 and NGC1277 (horizontal violet/green line). This value has to be compared with $\rho_{\rm pl}$ and for $\rho_{\rm pl} > \rho_{\rm sp}$ the enhancement from the DM spike is maximal and the density plateau can not form anymore. This happens for  annihilation cross-sections below $10^{-27} \rm cm^3 s^{-1}$. The spike profile of CenA (red horizontal line) has a larger maximal value and the DM density plateau forms for cross-sections up to $\sim 10^{-33} \rm cm^3 s^{-1}$. The case of a dSph (Reticulum II) with a IMBH of mass $M_{\rm BH} = 10^4 M_\odot$ is shown in the bottom panel: since dwarf galaxies are dominated by DM, the density spike is larger than the SMBH case (yellow horizontal line) and the density plateau always forms unless $\sigmav \leq 10^{-37} \rm cm^3 s^{-1}$. The value of the minispike profile, Eq.~\ref{eq:rhopsike}, is not affected quantitatively by the choice of the mass of the central IMBH.

Having quantified all parameters that enter in the flux estimate, Eq.~\ref{eq:dflux}, we can now proceed and examine the detection prospects for  monochromatic neutrino lines as a function of the sneutrino mass (which also indicates roughly the neutrino energy passing into the detector), as illustrated in Fig.~\ref{fig:flux_msnu}. Let us first consider the observation of dSPhs without the presence of DM spikes (first two panels, top left). The dSphs in the Northern sky can produce up-going muon events in an IC-like detector located at the South Pole: we show the flux expected by Draco (cyan), Ursa Major II (blue) and the stack of all dSphs, as listed in Tab.~\ref{tab:dwarf}. The expected flux coming from the dSph stack could be in the reach of a detector with the same sensitivity as IC40 for point source, while the flux from Draco could only be measured by the IC-Gen2 like detector. The coannihilation mechanism acting at freeze-out has the drawback of suppressing the overall flux from sneutrino annihilations at the present epoch, as the viable sneutrino DM configurations exhibit such small values of $\sigmav$. The usual dependence of the flux on the DM mass as $1/m^2_{\rm DM}$ ameliorates slightly the prospects for detection of $60-100$ GeV sneutrinos. By playing the same game with the dSPhs in the Southern hemisphere, which can give rise to up-going muons in a water detector such as KM3NeT, we see that the main contribution comes from Reticulum II (the flux from the stack of dSPhs coincides with the flux produced by Reticulum II, gray points). Since there are only a few known dSphs in the Southern Sky, the flux is rather small and below the sensitivity of KM3NeT. It should be noted however that the $68\%$ CL J-factor uncertainty ranges from $6.71 \times 10^{19}$ GeV$^2$ cm$^{-5}$ to $1.14 \times 10^{21}$ GeV$^2$ cm$^{-5}$~\cite{Bonnivard:2015tta}, which implies an uncertainty in the flux of roughly a factor 2.

The inclusion of a minispike in Draco or Reticulum II ameliorates the prospects for detection considerably (first two panels, bottom left). A minispike due to an IMBH  of $10^4 M_\odot$ (dark brown) or $10^2 M_\odot$(light brown) increases the flux by 4 or two orders of magnitude respectively in both dSphs. Both scenarios can be probed with our ideal neutrino telescopes in the configuration of IC40 or Antares, while next generation detectors would investigate more deeply the MSSM+RN. Even though the minispike can enhance the predicted flux by 4 orders of magnitude, it still lies far away from the reach of the present IC detector in its real configuration, because the point source sensitivity for neutrinos below 1 TeV is of the order $E\,  \Phi_{\nu} \sim 10^{-4} \rm GeV cm^{-2} s^{-1}$. In the right bottom panel, we show the dependence of the flux on the thermally-averaged cross-section. In the absence of a minispike in the inner core of Draco, only cross-sections of the order $10^{-28} \rm cm^3 s^{-1}$ can be reached with the future IC-Gen2, as the flux scales  linearly with respect to $\sigmav$. The DM minispike introduces a nontrivial dependence on $\sigmav$; the current IC40 sensitivity might be able to reach values down to $10^{-34} \rm cm^3 s^{-1}$, that could not realistically be  probed otherwise, for certain SUSY parameters.

As expected, the prospect for detection of the flux coming from SMBHs (right panel on top, as labelled) is less promising than the dSph case. This is caused by two phenomena: 1) the SMBHs are much distant objects than the dSphs, and  2) the total flux depends only on the DM density spike and not anymore on the density plateau, hence the flux cannot take advantage of the $R_S^3/D^2$ enhancement (i.e. the fact that the SMBH is actually very massive), as already anticipated in Sec.~\ref{sec:spike}. We are not considering this case any further.

\subsection{Complementarity with other WIMP searches}\label{sec:complDM}
\textbf{Gamma rays from dSphs - The Draco example:\\}
The sneutrino DM considered in this work appears to be quite elusive as far as the indirect DM searches are concerned. Sneutrinos annihilating in dSPhs and SMBHs produce a secondary spectrum of gamma rays, mainly from the decay of $W^+W^-$ or $b\bar{b}$ and their subsequent fragmentation into pions. Even assuming a branching ratio 100\% into $W^+W^-$ or into $b\bar{b}$, the current Fermi limits from dSPhs cannot constrain any of our sampled points, as the 95\% CL upper limit in $\sigmav$ versus DM mass lies in the ballpark from $ 3 \times 10^{-26} \rm cm^3 s^{-1}$ to $10^{-25} \rm cm^3 s^{-1}$~\cite{Ackermann:2015zua} for $100-500$ GeV DM mass. Under the same assumptions for the branching ratios, the gamma ray flux produced by sneutrino DM is also compatible with the limits set by the DM spike in M87~\cite{Lacroix:2015lxa}.

In the left panel of Fig.~\ref{fig:flux_msnu2},  we illustrate the complementarity between future gamma ray and neutrino telescopes. In particular we consider the Cherenkov Telescope Array~\cite{Acharya:2013sxa} (CTA), a ground based telescope array sensitive to high energy gamma rays. CTA is already in construction and may start its  first scientific run around 2016 with a partial array. Its expected point source sensitivity is denoted by the blue dotted line.  We show the expected flux for the neutrino line and for the gamma rays (orange points) from Draco, in the absence of a spike. The expected differential gamma ray flux is computed with \texttt{micrOMEGAS\_4.1} for all our sample points. We have integrated it in the energy range from 500 MeV up to DM mass and convoluted it with a J-factor of $0.1^\circ$~\cite{Bonnivard:2015xpq}, which is the expected angular resolution of CTA. Clearly the intensity of the secondary gamma ray flux is similar to the monochromatic neutrino line flux. This is relevant in the sense that neutrino telescopes can be very competitive with next generation of gamma ray telescopes, if their energy resolution is improved down to 100 GeV energy scale. 

\noindent\textbf{DM Direct detection:\\}
In the right panel of  Fig.~\ref{fig:flux_msnu2}, we illustrate the complementarity with direct detection searches. The values of the SI elastic cross-section for xenon detectors are plotted versus the expected neutrino line flux. Independently of the astrophysical assumptions (with minispike, light brown, or without, cyan), the configurations that lead to the largest monochromatic neutrino fluxes are in the reach of XENON1T~\cite{Aprile:2012zx} after two years of running (vertical green line). XENON1T is the first ton-scale detector for DM direct searches currently under construction and is supposed to release the first data in 2017. Even though direct detection searches are insensitive to the presence of minispikes in dSphs, an interesting complementarity arises when looking at the brown sample, which includes a central BH, $M_{\rm BH} = 10^4 M_\odot$, in Draco.  Indeed parameter configurations which are below the XENON1T sensitivity and close to the neutrino background~\cite{Billard:2013qya} (red vertical line) can be probed with the ideal IC-Gen2 or KM3Net detectors in five years time.

The hypothesis of having sneutrinos as subdominant DM components might produce enhanced ${\rm d}N/{\rm d}E_{\nu_{\rm line}}$ with respect to values obtained in our sample. Indeed subdominant sneutrinos can have large mixing angles, as the LUX exclusion bound gets weakened by the factor $\xi \equiv \Omega h^2_{\snu_{\tau}} / \Omega h^2_{\rm Planck}$ (where $\Omega h^2_{\rm Planck}$ is the value measured by Planck for the relic density), {\it i.e.} $ \xi \sigma^{\rm SI}_{\rm Xe} $. However, it should be noted that the enhancement in the intensity of the line is compensated by a reduction in the flux by $\xi^2$, so the full gain in visibility for the line is not straightforward. 

\noindent \textbf{Complementarity with LHC searches:\\}
Another interesting complementarity arises when considering the BSM searches at the LHC. As pointed out in Sec.~\ref{sec:intro}, the neutrino lines considered in this work, concern only the sneutrino and neutralino sectors. For the continuum, the additional Higgs boson contributions matter. If the rest of the MSMM+RN spectrum is very heavy, the prospects of observing signatures of MSSM+RN scenario at the LHC decrease rapidly. Indirect detection constraints at this point will be extremely useful in order to  constrain or even identify the sneutrino DM scenario. 
%
\section{Conclusions}\label{sec:concl}
In this article, we have investigated the detection prospects for monochromatic neutrino lines. We take advantage  of having a line signature pointing at the DM mass, and of having neutrinos as final-state particles. Indeed, neutrinos are particularly clean signatures as they propagate freely and undisturbed from the source to the detector, contrary to gamma rays which might undergo absorption processes and may not leave the most central region of the galaxy e.g. close to the BH, especially in the case of massive galaxies.  

We have studied the detection prospects of such monochromatic neutrino lines in two distinct astrophysical systems with a particularly high DM density: 1. the dwarf spheroidal galaxies with and without a DM spike hypothesis, and 2. the DM spikes in SMBHs. These environments are amongst the best probes in order to search for DM and have already set very stringent constraints in terms of gamma ray fluxes from DM annihilation.  Remarkably, neutrino telescopes chafe the potential of doing at least as well for certain classes of supersymmetric DM models.

As a particle physics candidate, we considered MSSM+RN, a well motivated and viable supersymmetric model with sneutrino DM. The RH field is a necessary addition to the MSSM to provide both a successful DM candidate, a mostly RH sneutrino, and to generate a Dirac neutrino mass term. The neutrino lines in this model are a direct outcome of the annihilation of sneutrinos at tree level.  Dark matter in dwarf galaxies spiked by IMBH provides an incredibly powerful means of probing low annihilation cross-sections well below $10^{-26} \rm cm^3 s^{-1}$ that are otherwise inaccessible by any  future direct detection or collider experiment. For instance, this signature depends only on a few SUSY parameters and in the case of heavy SUSY mass spectra, potentially out of reach of LHC, such neutrino lines can be a complementary signature to direct searches at the LHC. 

Sneutrino DM turns out to be very elusive in terms of indirect detection. Sensitive neutrino detectors in ice or water are required to probe the monochromatic neutrino fluxes from  dwarf spheroidal galaxies. The ideal set-up would be a detector equivalent to IceCube but with increased granularity for the string and optical modules, in order to achieve the same resolution for $\mathcal{O}(300)$ GeV DM similar to those for ultra high energy neutrinos from point sources. For reasonable masses of the IMBH in the center of the dSphs, the expected flux of neutrinos is similar to the gamma ray flux. In this case, neutrino detectors would be competitive with gamma ray telescopes, such as CTA, and add  further complementarity to  the multiprobe DM search approach.

From a model building point of view there might be ways of increasing the tree level annihilation into neutrinos. Indeed for the Dirac neutrino mass term,  the neutrino Yukawa coupling is very small and hence the line is sensitive to the LH component only, thus there is no tight connection between the neutrino line and neutrino mixing. The mixing angle in the sneutrino sector is instead limited by constraints from the LUX direct detection experiment. It may however be different, for instance, for the low-scale seesaw mechanism or seesaw type-I, where the neutrino masses are not generated by small neutrino Yukawa couplings.  A more general investigation of such MSSM extensions and their impact on neutrino line detection might be interesting and is left for future work.  

The possibility of detecting monochromatic neutrino lines from DM annihilations in several nearby astrophysical systems provides an especially appealing case to look for a smoking gun signature  of a plausible dark matter candidate, the sneutrino, that is predicted by supersymmetry. The next five years of LHC running will be crucial for searching for supersymmetric signatures, and the complementary approach offered by future neutrino telescopes may be equally rewarding for exploring the sneutrino as a DM candidate, and provide an excellent opportunity for fleshing out the multi-messenger study of the nature of DM.

%
%
%
%
\section*{Acknowledgments}
CA would like to thank M. Cirelli for providing the fit for neutrino flavor oscillations through Earth and T. Lacroix for useful discussion on the DM spikes. SK would like to thank G. Belanger and A. Pukhov for help with micromegas and R. Zukanovich Funchal and J. Pradler for discussions on IceCube detector. The research of CA and JS has been supported at IAP by the ERC project  267117 (DARK) hosted by Universit\'e Pierre et Marie Curie - Paris 6, PI J. Silk. SK is supported by the ``New Frontiers'' program of the Austrian Academy of Sciences. JS acknowledges the support at JHU by NSF grant OIA-1124403 and by the Templeton Foundation.

\bibliographystyle{JHEP}
\bibliography{biblio}

\end{document}